\documentstyle[aps,pre,epsf,psfig]{revtex}
\begin{document}

\title{Fluctuations, response and aging dynamics
in a simple glass-forming liquid out of equilibrium.}
\author{Walter Kob}
\address{Institut f\"ur Physik, Johannes Gutenberg-Universit\"at,
Staudinger Weg 7, D-55099 Mainz, Germany}

\author{Jean-Louis Barrat}
\address{ D\'epartement de Physique des Mat\'eriaux \\ Universit\'e
Claude Bernard and CNRS, 69622 Villeurbanne Cedex, France}
\maketitle

\begin{abstract}
By means of molecular dynamics computer simulations we investigate the
out of equilibrium relaxation dynamics of a simple glass former, a
binary Lennard-Jones system, after a quench to low temperatures. We
find that one-time quantities, such as the energy or the structure
factor, show only a weak time dependence. By comparing the out of
equilibrium structure factor with equilibrium data we find evidence
that during the aging process the system remains in that part of phase
space that mode-coupling theory classifies as liquid like.  Two-times
correlation functions show a strong time and waiting time $t_w$
dependence.  For large $t_w$ and times corresponding to the early
$\beta$-relaxation regime the correlators approach the Edwards-Anderson
value by means of a power-law in time. For large but fixed values of
$t_w$ the relaxation dynamics in the $\beta$-relaxation regime seems to
be independent of the observable and temperature. The
$\alpha$-relaxation shows a power-law dependence on time with an
exponent which is independent of $t_w$ but depends on the observable.
We find that at long times $\tau$ the correlation functions can be
expressed as $C_{\rm AG}(h(t_w+\tau)/h(t_w))$ and compute the function
$h(t)$. This function is found to show a $t$-dependence which is a bit
stronger than a logarithm and to depend on the observable considered.
If the system is quenched to very low temperatures the relaxation
dynamics at long times shows fast drops as a function of time. We
relate these drops to relatively local rearangements in which part of
the sample relaxes its stress by a collective motion of 50-100
particles.  Finally we discuss our measurements of the time dependent
response function. We find that at long times the correlation functions
and the response are not related by the usual fluctuation dissipation
theorem but that this relation is similar to the one found for
spin glasses with one step replica symmetry breaking.

\end{abstract}


\noindent

\section{Introduction}

It is well known from experiments that the most prominent feature of
supercooled liquids and glass forming systems is the rapid increase of
their relaxation time as temperature is decreased~\cite{review_expts}.
Since the early eighties, computer simulations have become a powerful
tool to investigate  such systems and hence
to increase our understanding of their extraordinary dynamical
behavior~\cite{review_sim}. However, because of the rapid increase of
the relaxation times these simulations are faced with the intrinsic
problem that, over a large range of temperature, systems cannot be
equilibrated, as the simulation time, typically $10^{-7}$s for an
atomic system, is smaller than the relaxation time. This means that on
a computer, any system with a longer relaxation time can only be
studied under nonequilibrium conditions. In other words, in computer
simulations the ``glass transition'' always takes place at temperatures
higher than in real experiments.

Despite this drawback, it has been shown that computer simulations can
provide useful information on the mechanisms than underly the rapid
increase of the time scales in supercooled liquids. One of the most
valuable contributions in this direction was the proof that the so
called mode-coupling theory~\cite{mct,mct_exp} does indeed provide a
quantitative description of this slowing down within the time window
explored in the simulation, at least for those simple atomic liquids
that are ``fragile glass formers'', i.e. show a strongly non-Arrhenius
dependence of the relaxation times as a function of temperature. The
findings from computer simulations \cite{review_sim} and experiments
\cite{mct_exp}  have largely confirmed that this theory provides a
consistent picture for the dynamical behaviour of such liquids over a
relaxation time range that covers several decades, typically
$10^{-11}-10^{-8}$s.

For larger relaxation times (lower temperatures), equilibrium
simulations are no longer possible.  On the other hand, one can take
advantage of the flexibility of the simulation to explore the {\it
nonequilibrium} properties  on times scales that would be  rather
difficult to access in experiments.  This possibility stems from the
fact that the simulations allows instantaneous quenches to low
temperatures, and a monitoring of the subsequent evolution on a fast
time scale. Experimentally, the study of nonequilibrium evolution  in
glassy systems following a fast quench is a field that has been
explored since a long time in the so called ``aging'' experiments
\cite{struik}, performed on a time scale of hours or even days.  More
specifically, ``physical aging'' consists in a slow evolution of the
characteristic properties of glassy systems that do not undergo any
changes in their chemical composition. Of particular interest is the
evolution of response functions (typically the elastic compliance) that
are usually found, e.g. in polymeric systems, to exhibit a strong
slowing down as the {\it waiting time} (i.e. the time elapsed since the
quench)  increases. The recent interest of the physics
community  in this behavior 
was aroused when a number of experiments performed on spin
glasses showed that a very similar behavior could be observed in these
disordered magnetic systems. In particular it was found that the
response to a magnetic field of a system that has been quenched into
its low temperature, nonequilibrium spin glass phase, becomes more and
more sluggish as the waiting time increases \cite{spin_glass_expts}.
Subsequently, a number of theories were put forward  to explain this
aging behavior of spin glass systems from first principles statistical
physics. A comprehensive review of these theories can be found in
reference \cite{review_aging_theory}.  Essentially, one is lead to
distinguish between phenomenological ``trap models'', domain growth
theories, and mode-coupling or mean field theories. ``Trap models''
describe the evolution of systems as a random walk in a complex phase
space, which can under certain conditions give rise to aging. The
domain growth models assume that the aging results from a coarsening
process somewhat similar to what can be observed in a ferromagnetic
system quenched below its critical temperature.  Mean field or
mode-coupling theories account exactly for the aging behaviour of some
disordered models in the limit of high dimension (e.g. a particle in a
random potential in a high dimensional space).  Interestingly, these
mean field theories give rise to a mathematical structure that is very
similar to that of the mode-coupling theory commonly used to describe
the dynamical properties of supercooled liquids~\cite{mct,mct_exp}.
  A major difference,
however, is that high dimensional systems can be shown to have a
true transition towards a non-ergodic behavior as the temperature is
lowered, whereas it is well known that the transition predicted by the
mode-coupling theory of liquids is smeared out and does not give rise
to a true singularity.  If, however, a true transition is present, a
natural consequence is that a system quenched below the transition
temperature will display aging behavior.  Theoretically, this aging
behavior of mean field systems was studied in great detail and is now
quite well understood \cite{review_aging_theory}. The question arises,
however, to what extent this description is relevant to the aging
behaviour of real, three dimensional systems.  In particular, in view
of the mathematical similarity mentioned above, it is quite natural to
inquire whether the nonequilibrium dynamics of those supercooled
liquids that are known to be well described by mode-coupling theory can
be accounted for by the mean field description of aging.

In an attempt to partially answer this question, we present molecular
dynamics simulations of a simple glass forming liquid, whose properties
under equilibrium conditions have been extensively studied in previous
work~\cite{kob_lj}.  We will concentrate on the
characterization of the ``aging'' behaviour of such a system on the time
scales that can be investigated in molecular dynamics simulations, i.e.
a few nanoseconds.  Once again, we emphasize  that these time scales
are very different from those usually investigated in aging experiments
on glasses  (although some recent dielectric spectroscopy experiments
\cite{leheny,israeloff} investigate relatively high frequency
behaviour). Moreover, the systems under study would, from an
experimental viewpoint, be considered as supercooled liquids rather
than glasses. Glassy behavior is observed only
 on the restricted time scale of the computer simulation. The
possibility of exploring such relatively short time scales is
nevertheless interesting, as it is precisely on such time scales that
mode-coupling theory is successful.  Moreover, the simulation offers
the opportunity to simultaneously compute time dependent correlation
functions, static quantities and response functions. Relating these
quantities to each other is an essential achievement in mean field
theories of aging behaviour, whose predictions can then be tested in
detail.

The paper is organized as follows. In section II, we recall the main
features of our model and describe some technical details of the
simulations.  Sections III and IV describe the aging behavior of static
properties and of correlation functions, respectively.
 Section V deals with the aging
behaviour of response functions.  Preliminary accounts of these results
can be found in \cite{kob97}. Some closely related studies, on a
slightly different system, have been reported in \cite{parisi97}.

\section{Model and Details of the Simulation}

The model we use is a binary (80:20) mixture of particles, which in the
following we will call A and B particles, that interact via a
Lennard-Jones potential, $V_{\alpha\beta}(r)=4\epsilon_{\alpha \beta}
\{(\sigma_{\alpha\beta}/r)^{12}-(\sigma_{\alpha\beta}/r)^6\}$, with
$\alpha, \beta \in \{{\rm A,B}\}$. The constants
$\epsilon_{\alpha\beta}$ and $\sigma_{\alpha\beta}$ are given by
$\epsilon_{\rm AA}=1.0$, $\sigma_{\rm AA}=1.0$, $\epsilon_{\rm
AB}=1.5$, $\sigma_{\rm AB}=0.8$, $\epsilon_{\rm BB}=0.5$, and
$\sigma_{\rm BB}=0.88$ and the potential is cut off and shifted at a
distance $2.5\sigma_{\alpha\beta}$. In the following we will report the
results in reduced units, with $\sigma_{\rm AA}$ and $\epsilon_{\rm
AA}$ the unit of length and energy, respectively (setting the Boltzmann
constant $k_B=1.0$). Time will be measured in units of
$\sqrt{\sigma_{\rm AA}^2m/48\epsilon_{\rm AA}}$, where $m$ is the mass
of the particles.

The simulation was done with $N=1000$ particles and at constant volume
$V=L^3$, with a box length $L=9.4$. At this density the {\it
equilibrium} dynamics of the system has been investigated
intensively~\cite{kob_lj,gleim98} and it has been found that at low
temperatures, $0.446 \leq T \leq 0.8$, this dynamics is described very
well by mode-coupling theory~\cite{mct,mct_exp} with a critical temperature
$T_c=0.435$.  Although there is evidence that for temperatures $T\leq
0.452$ the increase of the relaxation times, or the inverse of the
diffusion constant, is less strong than the power-law predicted by MCT,
$(T-T_c)^{-\gamma}$, with $\gamma=2.3$~\cite{nauroth97}, the increase is
still so strong that at temperatures less than $T_c$ the relaxation
time of the system exceeds by orders of magnitude the time scale
accessible to a present state of the art computer simulation ($O(10^7)$
time units). Hence for a computer simulation $T_c$ plays the role of
the glass transition temperature $T_g$.

The equations of motion were integrated with the velocity form of the
Verlet algorithm, with a step size of 0.02. The system was equilibrated
in the high temperature phase at a temperature $T_i>T_c$ and then
quenched at time $t=0$ to a temperature $T_f\leq T_c$. This quench was done
by means of a stochastic heat bath, i.e. every 50 time steps we
replaced the velocities of all the particles with ones that were
drawn from a Maxwell-Boltzmann distribution corresponding to a
temperature $T_f$. In order to study the dependence of the relaxation
behavior on the initial and final temperature we considered several
values of $T_i$ and $T_f$:  $T_i \in \{5.0, 0.8, 0.466\}$ and $T_f \in
\{0.435, 0.4, 0.3, 0.1\}$. 

After the quench we propagated the system for a waiting time $t_w$
after which the measurements of the quantities of interest were
started. In the course of the simulations we realized that there are
appreciable sample to sample  variations of the relaxation curves.
Therefore it was necessary to average  for each combination of $T_i$
and $T_f$ over 8-10 independent runs.

\section{One-time quantities}

One-time quantities are observables which in equilibrium are constants,
such as the total energy of the system, the pressure or, in a magnetic
system, the magnetization. In the non-equilibrium situation these
observables depend on the time which has elapsed since the quench and
thus are commonly called ``one-time'' quantities.

\begin{figure}[h]
\psfig{file=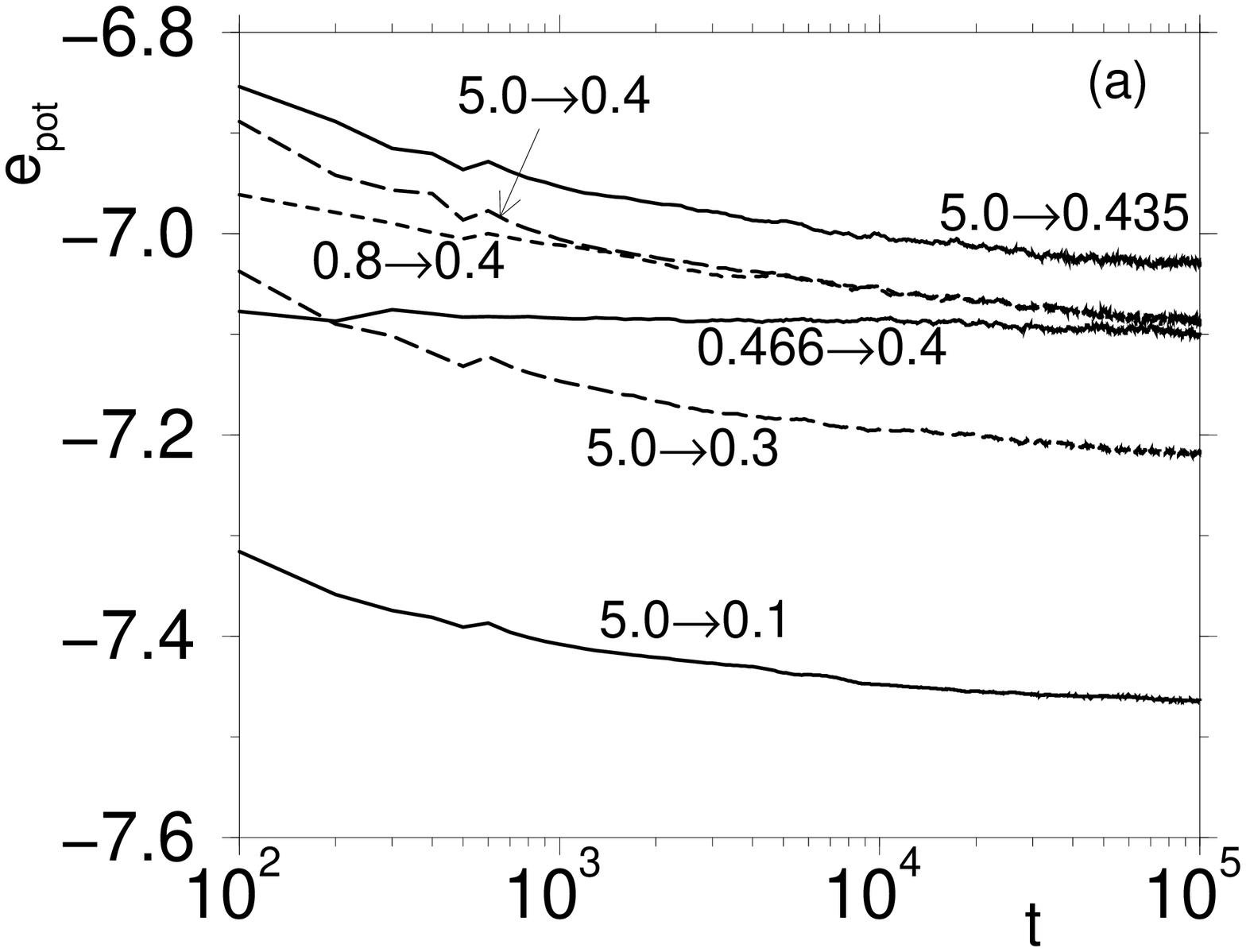,width=13cm,height=9.5cm}
\psfig{file=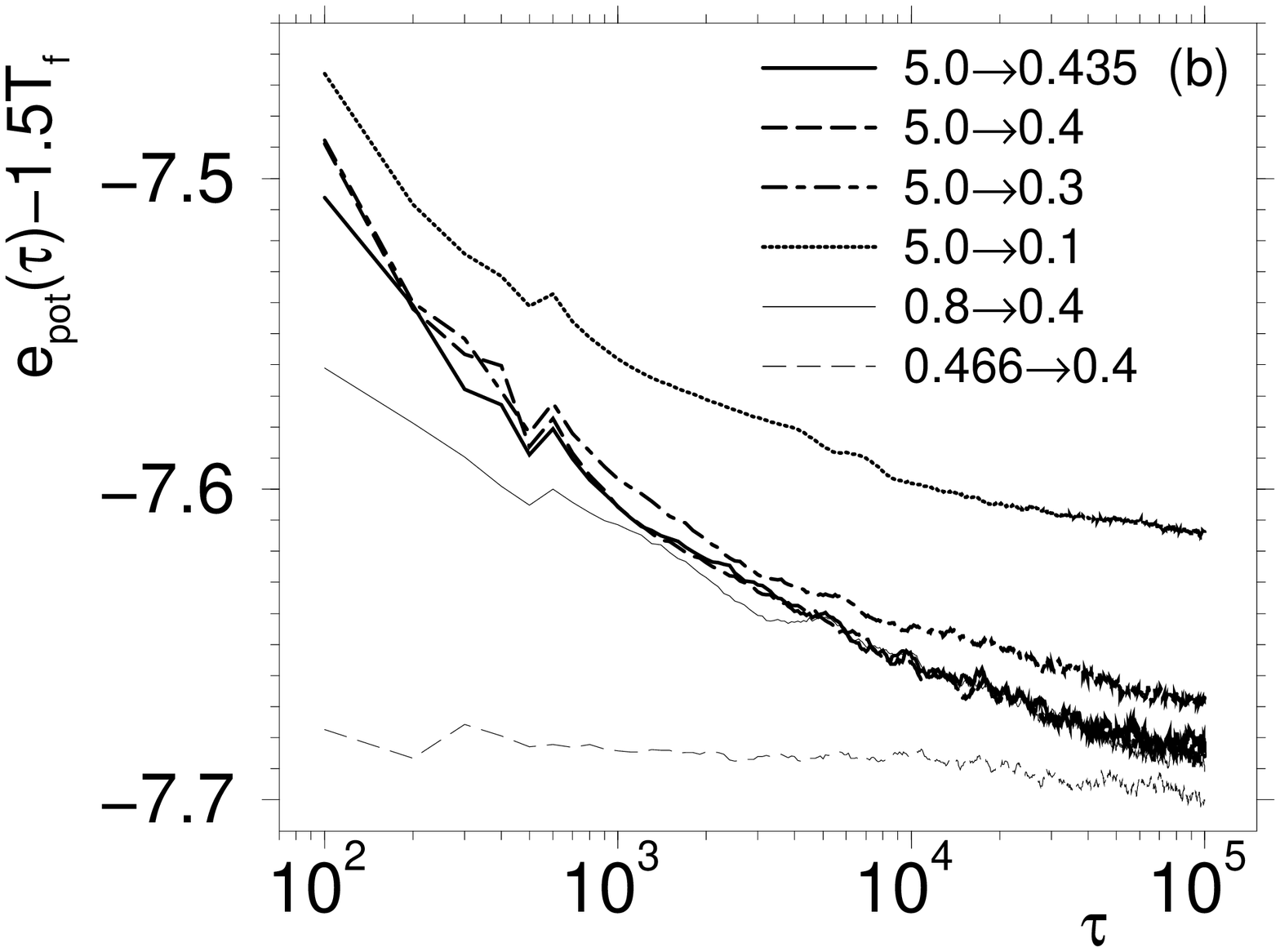,width=13cm,height=9.5cm}

\caption{a) Time dependence of the potential energy for various
combinations of $T_i$ and $T_f$. b) The same date as in a) but
shifted by 3/2$T_f$.}
\label{fig1}
\end{figure}

It has been shown before, see e.g.
Refs.~\cite{kob97,andrejew96,parisi97}, that the time
dependence of such quantities is relatively weak. In Fig.~\ref{fig1} we
show the time dependence of $e_{\rm pot}(t)$, the potential energy per
particle, for various combinations of $T_i$ and $T_f$. From the figure
it becomes evident that this time dependence is rather weak. In
Ref.~\cite{kob97} we have shown that it can be approximated well by a
power-law with a small exponent (0.14), or alternatively by a
logarithmic time dependence. Qualitatively this result agrees with the
one of Monte Carlo simulations of a soft sphere
system~\cite{parisi97} and a polymer model~\cite{andrejew96}
in which also power-law dependences have been observed for one-time
quantities. However, the exponents found in these simulations (0.7 and
1/3) are significantly larger than our, thus indicating that they are
not universal quantities. The large exponent of
reference~\cite{parisi97} might, however, be also due to the fact that
in that simulation $T_i$ was infinity and that the runs were rather
short, i.e. that the dependence of $e_{\rm pot}$ on time was not the
one valid for very long times.

From Fig.~\ref{fig1}a we conclude that at long times the curves do not
depend significantly on the value of the initial temperature (see the
curves with $T_f=0.4$). For short times, however, such a dependence can
be seen, in that, e.g., the curve with $T_i=0.466$ is almost constant
and shows only at long times a time dependence which is similar to the
one for higher values of $T_i$ (see also Ref.~\cite{kob97}). This time
dependence starts to be observable only for times that are comparable
with the $\alpha$-relaxation time of the system at $T_i$, which for
$T_i=0.466$ is on the order of $10^4$ time units~\cite{kob_lj}, since
this is the typical time scale that the system needs to find a
significantly different configuration. This estimate is, however,
only a lower bound, since the dynamics of the system is given by the
temperature $T_f<T_i$, and is hence slower than the one at $T_i$. See
also Ref.~\cite{age_is} for a further discussion of this point.

The figure also shows that there is a significant dependence of the
curves on $T_f$ in that the values of $e_{\rm pot}(t)$ decrease with
decreasing $T_f$. In a harmonic solid at a temperature $T_f$ one
expects that the (equilibrium!) potential energy varies like
$3/2k_BT_f$ and thus it is reasonable to subtract this ``trivial
contribution'' from the curves.  The result of this subtraction is
shown in Fig.~\ref{fig1}b from which we see that this procedure does
indeed make the curves collapse reasonably well. The most significant
exceptions are the curves ($T_i=0.466$, $T_f=0.4$) and ($T_i=5.0$,
$T_f=0.1$). These deviations can be understood by realizing that during
the aging process the system is looking for configurations with low
lying energies. If this search is started at a relatively low
temperature, such as $T_i=0.466$, the system will typically be able to
find configurations that have a lower potential energy than in the case
when $T_i$ is large thus explaining why the curve for $T_i=0.466$ is
below the other ones. In addition to this, such a search will be more
efficient at a higher temperature $T_f$ than at a low one, since then
the system has a better chance to overcome small local barriers. This
explains why the curve for ($T_i=5.0$, $T_f=0.1$) is above the other
ones with the same value of $T_i$.

\begin{figure}[h]
\psfig{file=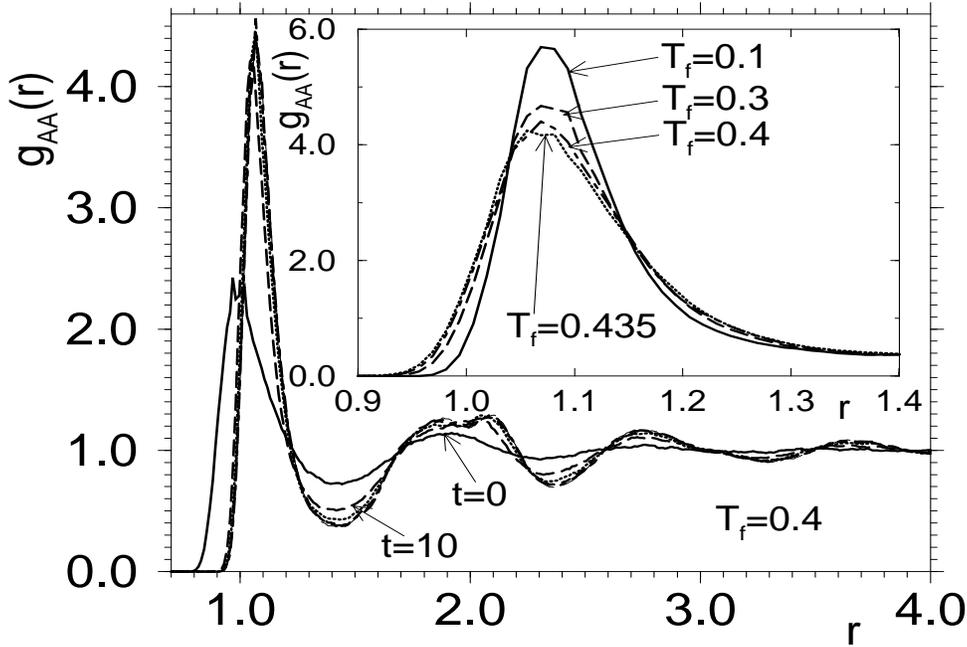,width=13cm,height=9.5cm}
\caption{Main figure: radial distribution function $g_{\rm AA}(r)$ for
times $t=0$ (before the quench) and $t=10$, 100, 1000, 10000, and
63100. $T_i=5.0$, $T_f=0.4$. Inset: $g_{\rm AA}(r)$ at long times for
$T_i=5.0$ and different values of $T_f$.}
\label{fig2}
\end{figure}

Since also the radial distribution function $g(r)$ is a one-time
quantity we expect that also its dependence on time is weak. That this
is indeed the case is shown in the main part of Fig.~\ref{fig2} where
we show $g_{\rm AA}(r)$, i.e. the partial radial distribution function
for the A particles (see Ref.~\cite{hansen86} for its definition) for
the times $t=0$ (i.e. before the quench), $t=10$, 100, 1000, 10000, and
63100 time units. From the figure we see that immediately after the
quench $g_{\rm AA}(r)$ changes its shape very quickly in that, e.g.,
the first nearest neighbor peak which in the high temperature phase
(curve labeled with $t=0$) is not very high, becomes much higher and
narrower.  For times larger than 10, the shape becomes essentially
independent of time, as expected for a one-time quantity and in
agreement with the results of Ref.~\cite{parisi97}.

\begin{figure}[h]
\psfig{file=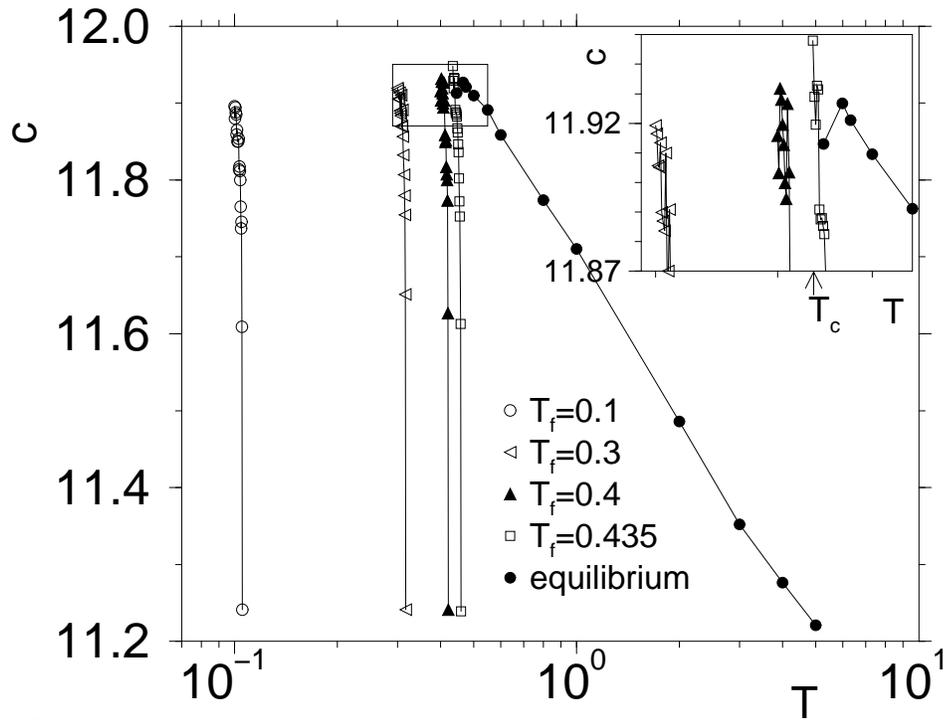,width=13cm,height=9.5cm}
\caption{Main figure: $T_f$ and time dependence of the area under the
first peak in $g_{\rm AA}(r)$ (open symbols). Closed symbols: Temperature
dependence of this quantity in equilibrium. Inset: Enlargement of the
region maked by a box in the main figure.}
\label{fig3}
\end{figure}

That the form of $g_{\rm AA}(r)$ {\it at long times} has a significant
dependence on $T_f$ is demonstrated in the inset of Fig.~\ref{fig2},
where we show the main peak for different values of $T_f$ at long times
($t=63100$). As expected we find that the height of this peak increases
with decreasing temperature and that it becomes narrower. Thus we
conclude that the typical configurations in which the system is stuck
after long times does depend on the final temperature. Very recently
Latz made the interesting proposition that the typical configurations
in which the system is stuck at long times during the aging process
share a common property in that all of them are very close to the
so-called critical surface of mode-coupling theory
(MCT)~\cite{mct,mct_exp} which divides the liquid like phase of the
system from its glass like phase~\cite{latz99}. MCT predicts that this
critical surface can be calculated by using as input the radial
distribution function, i.e. a purely {\it static}
quantity~\cite{mct,mct_exp}. Furthermore it has been shown that, for
simple liquids, the relevant part of this input is related to the area
under the first peak in the structure factor~\cite{mct,bengtzelius84},
or alternatively, to the area under the first peak in $g(r)$ weighted
with $4\pi r^2$. In order to test the validity of the proposition by
Latz we have therefore calculated the integral of $4\pi r^2g_{\rm
AA}(r)$ between zero and 1.406, the location of the first minimum in
$g_{\rm AA}(r)$. Note that this integral, which we call $c$, is the
partial coordination number. The time and $T_f$ dependence of $c$ are
shown in the main part of Fig.~\ref{fig3} (open symbols). (In order to
expand the axis at low temperatures we show this data versus the
logarithm of $T_f$.). The times $t$ shown are spaced essentially
equidistant on a logarithmic time axis and are 0, 10, 40, 60, 100, 160,
250, 400, 630, 1000, 1580, 2510, 3980, 6310, 10000, 15850, 25120,
39810, and 63100. (In order to show the time dependence we plot the
data for $t=63100$ at $T=T_f$ and with decreasing $t$ at a temperature
which is a factor of 1.003 higher than the previous data point.) From
this figure we recognize that with increasing time the value of $c$
increases rapidly and then becomes constant to within the precision of
our data. The value of this limiting constant, which we call
$c_{\infty}$, increases weakly with increasing $T_f$ but to a first
approximation it can be considered as independent of $T_f$, thus
supporting the prediction of Latz.

In order to see whether the $T_f$ dependence of $c_{\infty}$ is indeed
weak it is useful to compare it with the temperature dependence of the
area under the first peak $g_{\rm AA}$ {\it in equilibrium}. For this
we have included in the figure also the value of $c$ for temperatures
$5.0 \geq T \geq 0.446$ (filled symbols). (The $g_{\rm AA}(r)$ for the
various temperatures are results from other
simulations~\cite{kob_lj,gleim98}.) In this temperature range the
equilibrium value of $c$ shows an appreciable temperatures dependence
(which is close to a logarithmic dependence) which is indeed much more
pronounced than the $T_f$ dependence of $c_{\infty}$ of the
non-equilibrium simulations. To a first approximation the value of
$c_{\infty}$ is given by the value of $c$ of the equilibrium curve at
$T=T_c$, where $T_c$ is the so-called critical temperature of
MCT~\cite{latz99}. For the present system this critical temperature has
been estimated to be around $T=0.435$~\cite{kob_lj,gleim99}. Due to the
problem to equilibrate the system at temperatures close to $T_c$ it is
currently not possible to determine the equilibrium value of $c$ at
$T_c$ and therefore to compare this value with the $c_{\infty}$ from
the out-of-equilibrium simulations. However, since the equilibrium
value of $c$ is expected to show a smooth temperature dependence, it
can be estimated quite reliably from the equilibrium data at a bit
higher temperatures.  In the inset of Fig.~\ref{fig3} we show an
enlargement of the region which is marked by a box in the main figure
and which encloses the temperature range around $T_c$. From this inset
we see that the values of $c_{\infty}$ for $T_f=T_c=0.435$ and
$T_f=0.4$ are indeed very close to an extrapolation of the equilibrium
curve to $T_c$. Hence we conclude that it is indeed possible to
calculate from the equilibrium data with reasonable accuracy also
certain quantities for the out-of-equilibrium situation. This point
is discussed in more detail in Ref.~\cite{age_is}.

\section{Two-times quantities}

\subsection{General features of two time correlation functions}

In equilibrium, time correlation functions between any two observables
$A$ and $B$, $\langle A(\tau) B(0) \rangle$ depend only on the time
difference $\tau$, i.e. $\langle A(\tau) B(0) \rangle= \langle
A(\tau+t_w)B(t_w)\rangle$, for all waiting times $t_w$. For the out of
equilibrium situation this equality no longer holds and therefore
such time correlation functions depend on two quantities, the time
difference $\tau$ and $t_w$, the time passed since the generation of
the non-equilibrium situation. Therefore such correlation functions are
called two-times quantities and in this subsection we will demonstrate
that such quantities are very well suited to investigate the aging
dynamics of out of equilibrium systems.

For liquids in {\it equilibrium} the dynamics is often studied by means
of the intermediate scattering function $F(k,t)$ which is defined
by~\cite{hansen86}

\begin{equation}
F(k,\tau)=\frac{1}{N}\sum_{j,l} \exp(i{\bf k}\cdot 
({\bf r}_j(\tau)-{\bf r}_l(0))
\label{eq1}
\end{equation}

and by the its so-called self part, $F_s(k,t)$ given by

\begin{equation}
F_s(k,\tau)=\frac{1}{N}\sum_j \exp(i{\bf k}\cdot 
({\bf r}_j(\tau)-{\bf r}_j(0))\quad.
\label{eq2}
\end{equation}

Here ${\bf r}_k(\tau)$ are the positions of the particles at time $t$
and ${\bf k}$ is the wave vector. For the non-equilibrium situation
these functions are generalized to

\begin{equation}
C_{c,k}(t_w+\tau,t_w)=\frac{1}{N}\sum_{j,l} \exp(i{\bf k}\cdot ({\bf
r}_j(t_w+\tau)-{\bf r}_l(\tau))
\label{eq3}
\end{equation}

and

\begin{equation}
C_k(t_w+\tau,t_w)=\frac{1}{N}\sum_{j} \exp(i{\bf k}\cdot ({\bf
r}_j(t_w+\tau)-{\bf r}_j(\tau)).
\label{eq4}
\end{equation}

(Note that these last equations are trivially generalized to
multi-component systems.) For the present system the dependence of
$F(k,t)$ and $F_s(k,t)$ on time and temperature have been discussed
extensively in Refs.~\cite{kob_lj,gleim98}. In Fig.~\ref{fig4} we show
the time dependence of the corresponding quantities $C_k(t_w+\tau,t_w)$
and $C_{c,k}(t_w+\tau,t_w)$ for different waiting times $t_w$. The
values of $t_w$ are 0, i.e. immediately after the quench, and $t_w=10$,
100, 1000, 10000, and 63100 time units and the final temperature
$T_f=0.4$ (solid lines). In Fig.~\ref{fig4}a we show
$C_k(t_w+\tau,t_w)$ for $k=7.23$, the location of the first peak in the
partial structure factor of the A-A correlation~\cite{kob_lj}.  We see
that the correlation function decays quite quickly as a function of
$\tau$, which in view of the fact that we are at a very low temperature
might be surprising. However, one should realize that the decay of the
correlation function to zero does {\it not} imply that the system has
relaxed to equilibrium but only that the particle configuration at the
end of the measurement is uncorrelated with the configuration at the
start. From the figure we see that with increasing value of the waiting
time the decay of the curves occurs at longer and longer times. This
observation can be rationalized by realizing that the driving force
which leads to the decorrelation of the state at time $\tau$ from the
state at time zero decreases with increasing waiting time $t_w$ since
in the time between zero and $t_w$ the system had already the
possibility to relax. This $t_w$ dependence of the curves is, however,
only observed at long times. For short times the curves fall onto a
master curve and leave this master curve only at a time which increases
with $t_w$ (and is roughly proportional to $t_w$~\cite{kob97}). The
time regime in which this master curve is observed is usually called
``short time regime'' whereas the time regime in which the curves show
a significant $t_w$ dependence is called the ``aging regime''. In the
short time regime the particles rattle in the cages formed by their
nearest neighbors and this type of motion is thus not sensitive on the
value of $t_w$ (if $t_w$ is not too small). Only for longer times the
particles are able to leave this cage and the typical time scale for
this process depends strongly on $t_w$.

We also mention that the oscillation seen in the curves at $\tau=1.0$
originates from the coupling of the system to the heat bath. Since at
this time the velocities of all particles are swapped with the ones
drawn from a Maxwell-Boltzmann distribution at temperature $T_f$, the
motion of the particles is, on average, slowed down or even reversed.
Hence for a brief period the relaxation is slower than expected and
thus the correlation curve decays slower. This effect is, however, only
seen at short times and thus can be disregarded for large values of
$\tau$.

It is also interesting to compare the relaxation behavior of the system
in this non-equilibrium situation with the one in equilibrium. In
Fig.~\ref{fig4}a we have therefore also included the incoherent
intermediate scattering function $F_s(k,t)$ in equilibrium (dashed lines) 
which was obtained in a simulation of Gleim {\it et
al.}~\cite{gleim98}. The dashed line is a curve that corresponds to the
equilibrium relaxation dynamics of the system at $T=0.446$ (this was
the lowest temperature at which the system could be equilibrated).  For
very short times, $\tau\leq 1.0$, this curve essentially
 coincides with the above
discussed master curve in the short time regime. For times $\tau$
larger than 1.0 deviations are observed, which are {\it not} due to the
mentioned oscillations. A careful inspection of the curves shows that
the approach of the master curve and the dashed curve to the (quasi)
plateau at intermediate times is quite different from each other in
that the former shows a very slow approach whereas the one of the
latter is quite fast.

Also at long times we find differences in the relaxation behavior. In
equilibrium it has been demonstrated that the relaxation curves can be
fitted very well by the so-called Kohlrausch-Williams-Watts (KWW)
function, $A\exp(-(\tau/\tau_{\rm rel})^{\beta})$, where $\tau_{\rm
rel}$ is the relaxation time and an exponent $\beta \leq
1$~\cite{kob_lj}. From the figure we see that at long times the shape
of the equilibrium and non-equilibrium curves are quite different and a
more detailed analysis shows that a KWW function does not give a good
fit to the data. What seems to be the same, however, (or at least very
similar) is the height of the plateau, i.e. the so-called
Edwards-Anderson parameter~\cite{edwards75} or non-ergodicity
parameter.

In Fig.~\ref{fig4}b we show the relaxation curves for the collective
function, i.e. $C_{c,k}(t_w+\tau,\tau)$ for the same waiting times.
The wave-vector is now $k=9.60$, the location of the minimum in the
partial structure factor of the A-A correlation~\cite{kob_lj}.  Due
to the collective nature of this correlation function the statistics is
worse than the one for the self part but from the figure we see that
also this observable can be studied with satisfactory accuracy. Since
at this wave-vector the (equilibrium) value of the non-ergodicity
parameter has a local minimum~\cite{kob_lj} it is expected that also
the one for the non-equilibrium case is relatively small and from the
figure we see that this is indeed the case.

\subsection{Quantitative analysis of the relaxation at high temperatures}

In the following we will now analyze the relaxation behavior of the
system in more detail. For this we start first with the dynamics at
short times. Mean field calculations predict that for times at which
the correlators are in the vicinity of the plateau, a time regime which
in the equilibrium dynamics of glasses is called ``$\beta$-relaxation
regime, two power-laws are observed~\cite{review_aging_theory,LCPLD}.
 Denoting by $q_{\rm EA}$
the value of the Edwards-Anderson parameter, it is predicted that any
time correlation function $C(t_w+\tau,t_w)$ should behave like

\begin{equation}
C(t_w+\tau,t_w) \approx q_{\rm EA}+c_a \tau^{-a} \quad 
{\rm if } \quad C\geq q_{\rm EA}
\label{eq5}
\end{equation}

and

\begin{equation}
C(t_w+\tau,t_w) \approx q_{\rm EA}-c_b (\tau/\tau_{\alpha})^{b} 
\quad {\rm if } \quad C\leq q_{\rm EA} \qquad .
\label{eq6}
\end{equation}

Here $\tau_{\alpha}$ is the typical relaxation time at long times and
the exponents $a$ and $b$ are related by

\begin{equation}
m\frac{\Gamma^2(1+b)}{\Gamma(1+2b)}=\frac{\Gamma^2(1-a)}{\Gamma(1-2a)}.
\label{eq7}
\end{equation}

\begin{figure}[h]
\psfig{file=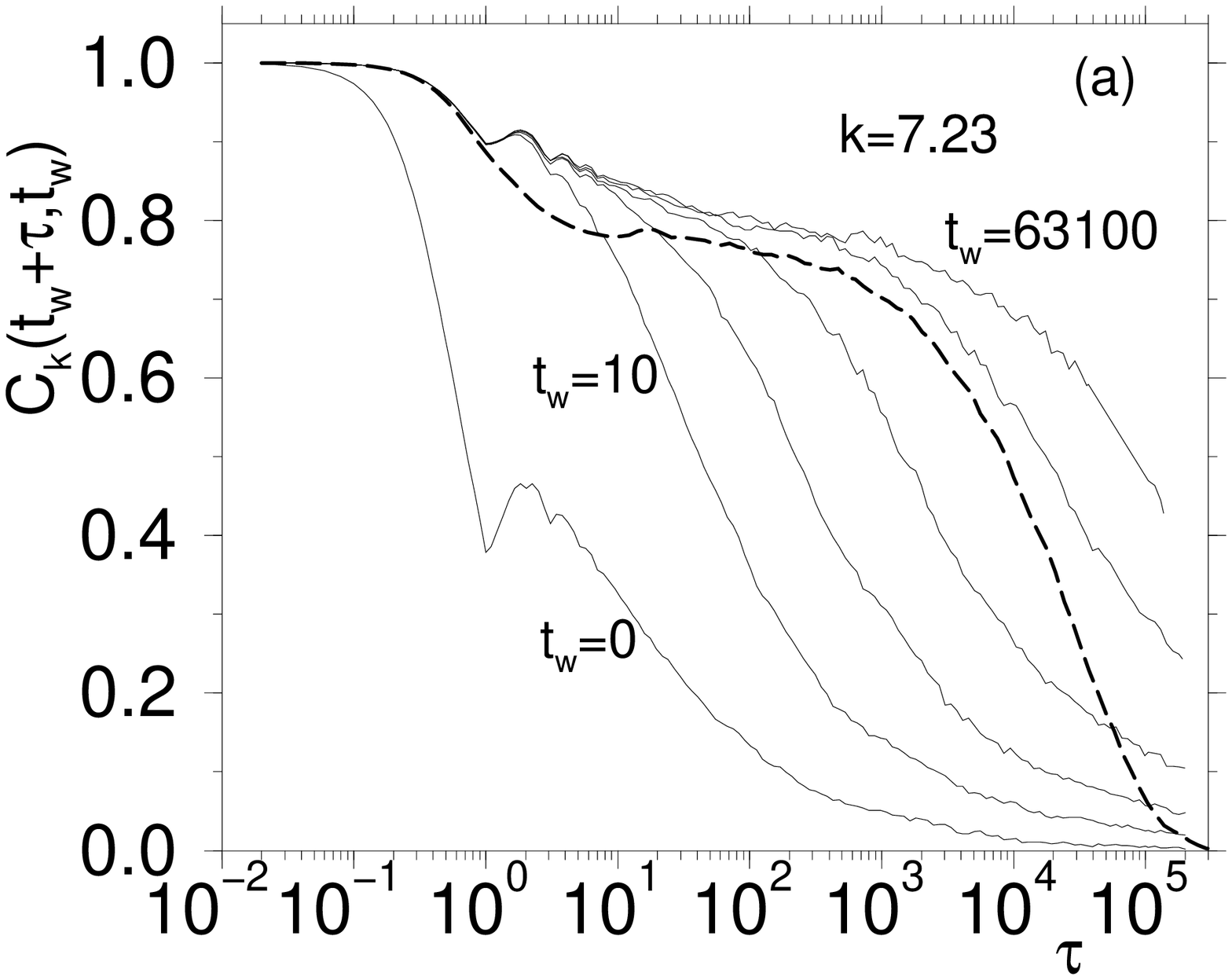,width=13cm,height=9.5cm}
\psfig{file=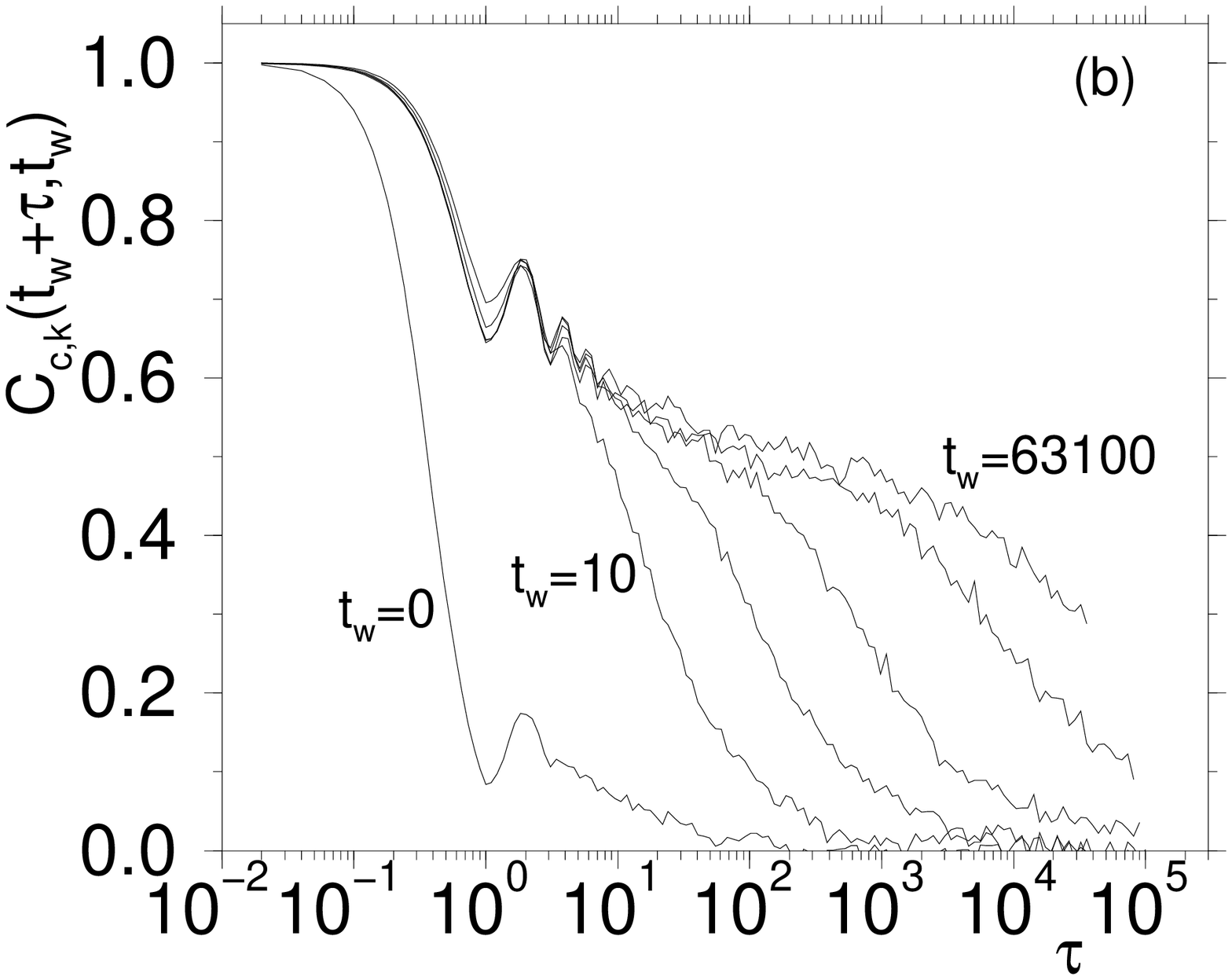,width=13cm,height=9.5cm}
\caption{Time dependence of the correlation functions $C_k(t_w+\tau,t_w)$,
pannel (a), and $C_{c,k}(t_w+\tau,t_w)$, pannel (b), for the waiting
times $t_w$=0, 10, 100, 1000, 10000, and 63100. In pannel (a) we have
also included an equilibrium curve at higher temperature (dashed
 line). See text for details.}
\label{fig4}
\end{figure}

\begin{figure}[h]
\psfig{file=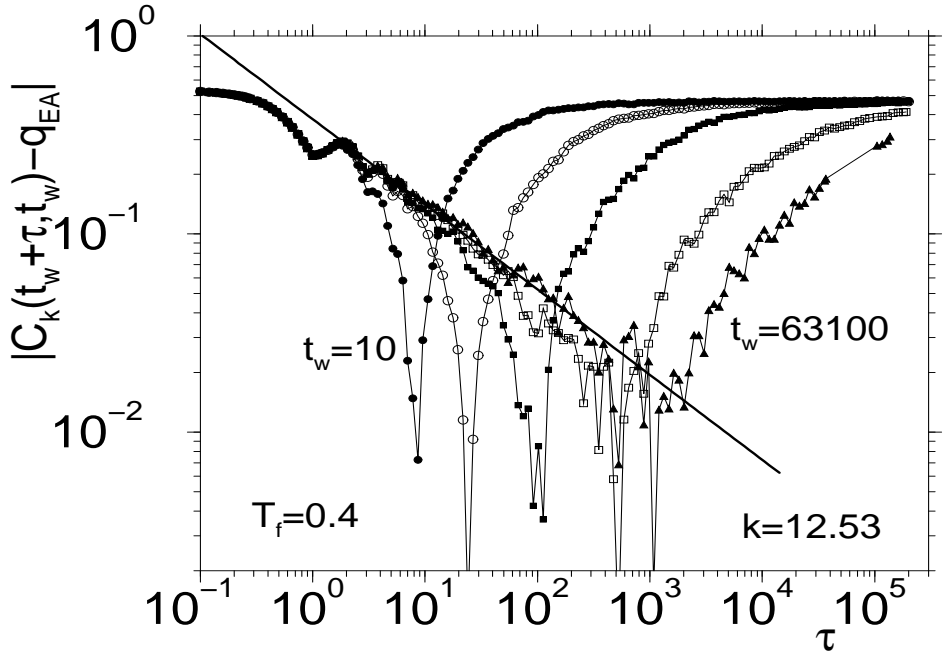,width=13cm,height=9.5cm}
\caption{Test of the presence of a power-law at short times in
$C_k(t_w+\tau,t_w)$. The value of $q_{\rm EA}$ is 0.47. $t_w$=10, 100,
1000, 10000, and 63100.  The bold straight line has slope 0.42.}
\label{fig5}
\end{figure}

The quantity $m$ in the last equation is the so-called
fluctuation-dissipation-violation factor and will be discussed later in
more detail. For the moment it is sufficient to know that it is
expected to be a system universal number equal or less than 1.0. We
also mention that the two power-laws of Eqs.~(\ref{eq5}) and
(\ref{eq6}) have been discussed for a long time and in great detail in
the MCT of supercooled liquids and glasses~\cite{mct,mct_exp}. In order
to see whether it is possible to see these power-laws in our data we
have estimated the value of $q_{\rm EA}$ for $C_k(t_w+\tau,\tau)$ for
$k=12.53$ and show in Fig.~\ref{fig5} $|C_k(t_w+\tau,\tau)-q_{\rm
EA}(k)|$ in a double logarithmic plot for different waiting times. This
value of $k$ is relatively large so that the value of $q_{\rm EA}$ is
relatively small, 0.47. A small value of $q_{\rm EA}$ is useful if one
wants to check for the presence of the power-law given in
Eq.~(\ref{eq5}) whereas for the check of Eq.~(\ref{eq6}) a large value
of $q_{\rm EA}$ is better.  (These conclusions are reached from the
analysis of equilibrium dynamics~\cite{mct_corrections} but it can be
expected that they hold also for the nonequilibrium case.) From the
figure we recognize that it is indeed possible to see a power-law in
the short time regime with an exponent around $0.42\pm 0.05$ (bold
solid line). In the discussion of Fig.~\ref{fig4} we said that in the
late part of the $\beta$-relaxation regime the equilibrium curves and
the non-equilibrium curves show a very similar time dependence and that
the one of equilibrium is given by a power-law. The exponent of this
power-law is around 0.63~\cite{nauroth97,gleim99}. If we assume the
relation (\ref{eq7}) to hold true one therefore obtains $m \approx
0.57$. Due to the relatively large error of the exponents $a$ and $b$
this number also has a significant error.

We have also checked for the presence of the power-law for
other values of $k$, as well as for the collective correlation
function $C_{c,k}(t_w+\tau,t_w)$ and found that all of them show such a
time dependence with an exponent which is compatible with 0.42.
Unfortunately, however, the exact determination of the exponent is
rather difficult, since a change of the (unknown) value of $q_{\rm
EA}$ will lead to a change in the exponents also. In order to avoid
this problem to some extend we use a trick which has proved to be
useful in the context of the analysis of equilibrium data (see, e.g.,
\cite{signorini90,horbach99}). For the equilibrium case MCT predicts
that {\it in the $\beta$-relaxation regime} the so-called factorization
property holds~\cite{mct,mct_exp}. This means that any time correlation
function $\phi(t)$ can be written as $\phi(t)=q_{\rm EA}(\phi)+h_{\phi}
G(t)$, where $h_{\phi}$ is a constant and the whole time dependence is
given by the $\phi$-independent, i.e. {\it system universal}, function
$G(t)$. If we assume that a similar relation holds also in the
non-equilibrium situation we have

\begin{equation}
\phi(t_w+\tau,t_w)=q_{\rm EA}(\phi)+h_{\phi}G(t_w+\tau,\tau)\quad .
\label{eq8}
\end{equation}

\begin{figure}[h]
\psfig{file=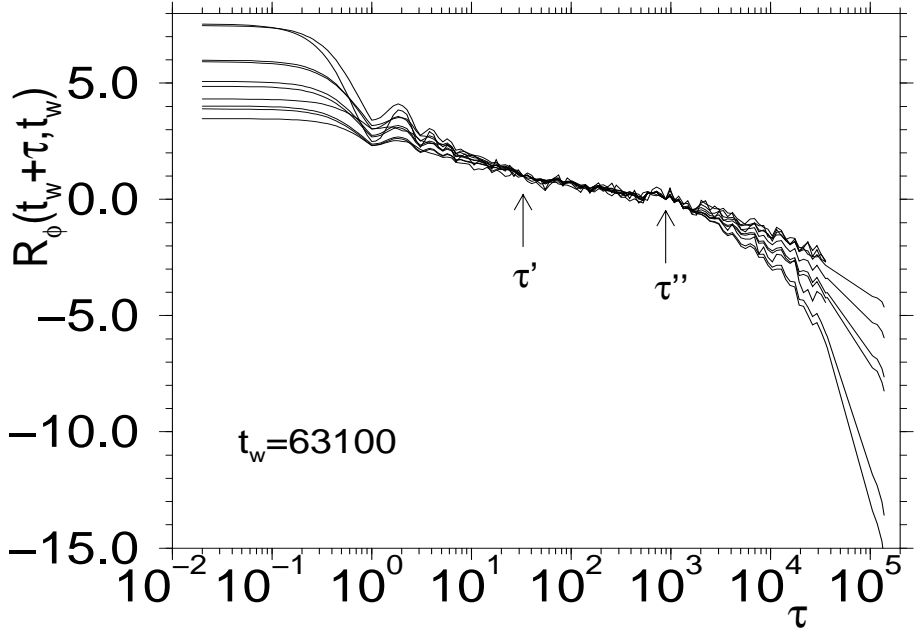,width=13cm,height=9.5cm}
\caption{Time dependence of the ratio $R_{\phi}(t_w+\tau,t_w)$ for
different correlation functions (see text). The vertical arrows show
the values of the times $\tau'$ and $\tau''$ used to calculate
$R_{\phi}$.}
\label{fig6}
\end{figure}

We see that if the exponents $a$ and $b$ as well as the quantity $m$ 
are independent of the observable that then Eqs.~(\ref{eq5}) and
(\ref{eq6}) are indeed compatible with such an Ansatz. From
Eq.~(\ref{eq8}) it follows immediately that if $\tau'$ and $\tau''$
are two arbitrary times in the $\beta$-relaxation regime, the ratio

\begin{equation}
R_\phi(t_w+\tau,t_w) =\frac{\phi(t_w+\tau,t_w)-\phi(t_w+\tau'',t_w)}{
\phi(t_w+\tau',t_w)-\phi(t_w+\tau'',t_w)}
\label{eq9}
\end{equation}

is independent of $\phi$ if $\tau$ is in the $\beta$-relaxation regime
also. In Fig.~\ref{fig6} we plot this ratio for different choices of
$\phi$. These are $C_{c,k}(t_w+\tau,t_w)$ for $k=6.52$, 7.23, 9.6, and
12.53 as well as $C_k(t_w+\tau,t_w)$ for the same wave-vectors and also
$k=2.0$ and $k=3.0$. The value of $t_w$ is kept fixed at $t_w=63100$.
From the figure we see that in the $\beta$-relaxation regime the 
different curves collapse nicely onto a master curve, thus
demonstrating the validity of the factorization property. That this
result is by no means trivial can be recognized from the fact that for
very short and very long times the different curves do not fall onto a
master curve at all. From the existence of the master curve in the
$\beta$-relaxation regime we thus come to the conclusion that, within
the accuracy of our data, in this time regime the time dependence of
the relaxation is governed by one single system universal function
$G(t_w+\tau,t_w)$. From the results shown in Fig.~\ref{fig5} we see that this function is compatible with power-law of
the form given by Eq.~(\ref{eq5}).

As we will see later, it is expected that the quantity $m$ depends on
the value of $T_f$~\cite{review_aging_theory}. Therefore one might conclude that also
the exponent $a$ and $b$ depend on temperature. From making fits to the
master curves in the $\beta$-relaxation regime of our data, see
Fig.~\ref{fig5}, it is hard to conclude whether or not such a
dependence is indeed present since the choice of the non-ergodicity
parameter affects the values of the exponents also. Therefore we have
calculated the ratios $R_{\phi}(t_w+\tau,t_w)$ also for lower values of
$T_f$, $T_f=0.3$ and 0.1, and found that the curves for the
different observables do indeed fall onto a master curve also. More
important is the observation that within the accuracy of our data this
master curve does not depend on $T_f$, thus giving evidence that the
function $G(t_w+\tau,t_w)$ is only a weak function of $T_f$.

We also mention that mean-field calculations lead to the expectation
that the quantities $a$, $b$ and $m$ in Eq.~(\ref{eq7}) do depend on
the final temperature. (This is in contrast to the equilibrium MCT
where $m=1$ and $a$ and $b$ are independent of temperature.) In
Sec.~\ref{secVC} we will show that $m$ shows a significant dependence
on the final temperature in that it becomes smaller with decreasing
$T_f$. Thus that result might seem to be in contradiction with the fact
that $G(t_w+\tau,t_w)$ seems to depend only weakly on $T_f$. The
resolution of this apparent problem is that if we assume $b$ to be
constant that a decrease of $m$ can be compensated by an increase of
$a$. It is, however, simple to see that if $m$ is small $a$ has to be
close to 0.5 and that a variation of $m$ can be compensated by a very
small change in $a$, i.e. without leading to an appreciable change in
$G(t_w+\tau,t_w)$.

\begin{figure}[h]
\psfig{file=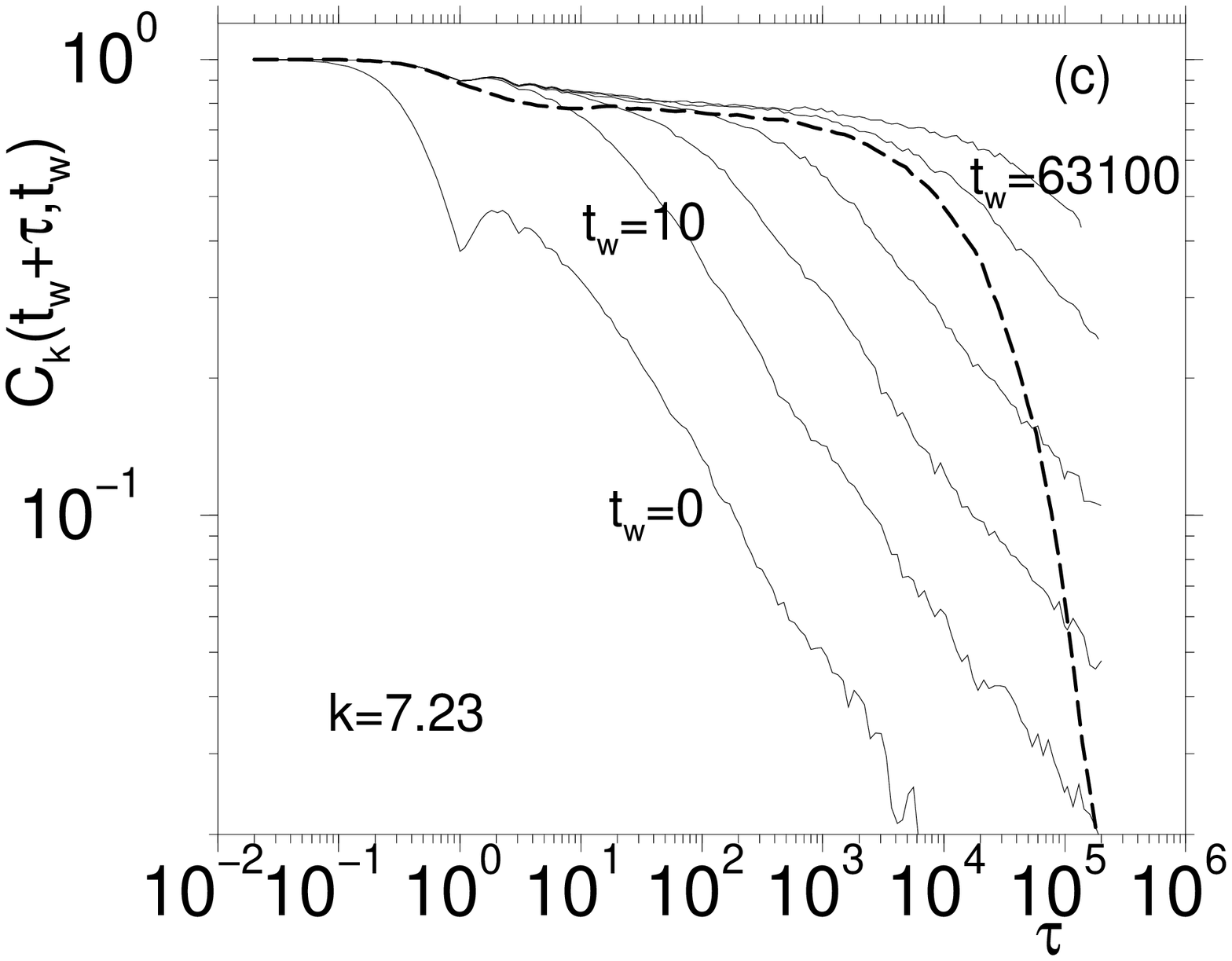,width=13cm,height=9.5cm}
\psfig{file=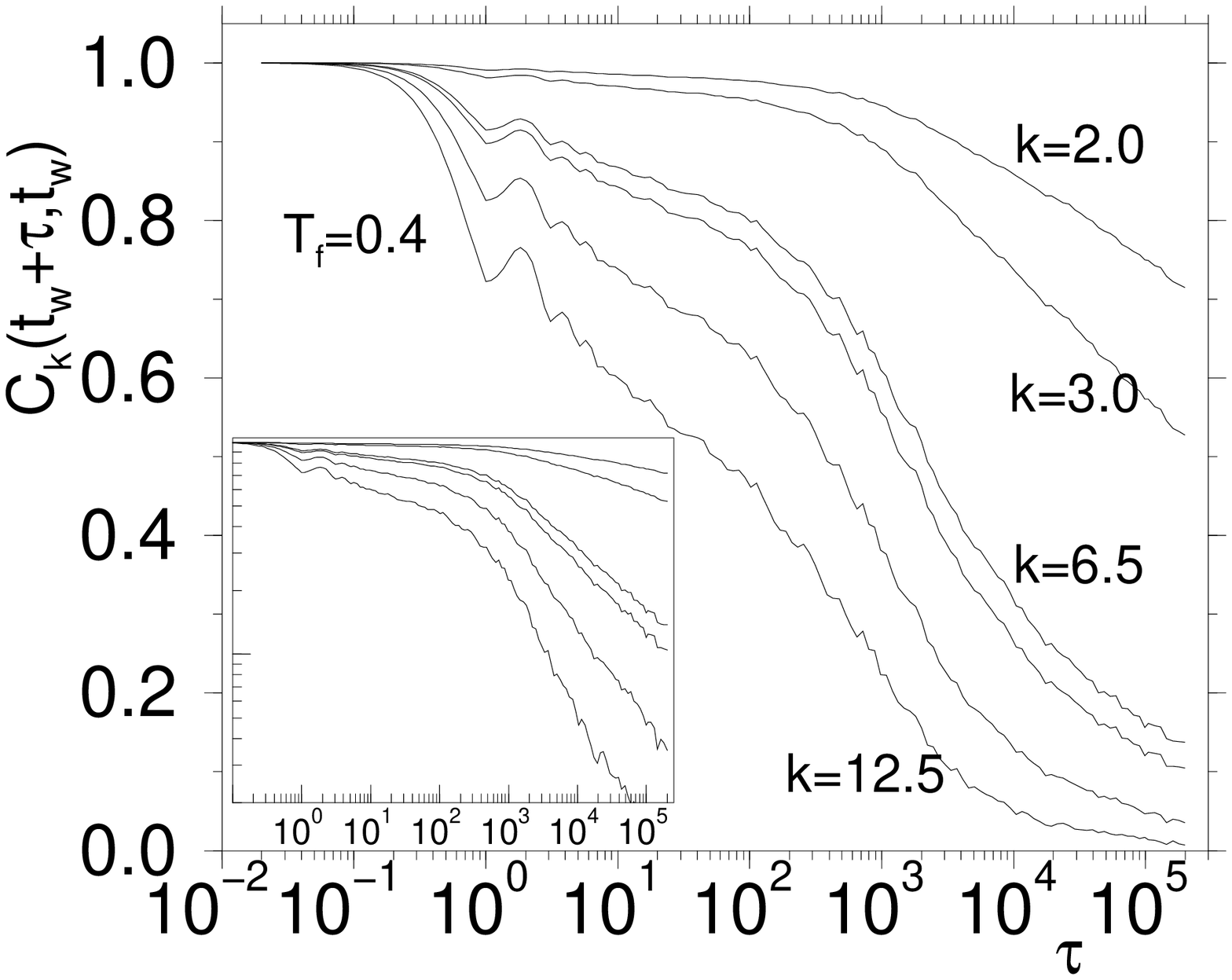,width=13cm,height=9.5cm}
\caption{a) Double logarithmic plot of $C_k(t_w+\tau,t_w)$ for the aging
dynamics (solid lines) and the equilibrium dynamics (dashed curve).
 The waiting times are: $t_w$=0, 10, 100, 1000, 10000,
and 63100. b) Wave-vector dependence of $C_k(t_w+\tau,t_w)$ for the
waiting time $t_w=1000$. From top to bottom the values of $k$ are
2.0, 3.0, 6.5, 7.23, 9.6, and 12.5. Inset: the same curves in a
double logarithmic presentation.}
\label{fig7}
\end{figure}

Having analyzed the relaxation behavior of the system in the
$\beta$-relaxation regime we now investigate the one of the
$\alpha$-regime, i.e. the relaxation of the system at long times.  When
we discussed the time correlation functions in Fig.~\ref{fig4}a we
already mentioned that in this regime the relaxation differs from the
one in the equilibrium situation in that the time correlation functions
can not be described well by the KWW-law. In Fig.~\ref{fig7}a we show
the same correlation functions as in Fig.~\ref{fig4}a, but this time in
a double log plot. From this presentation of the data we see that at
long times the out-of-equilibrium curves show a power-law. The exponent
is around 0.4 and is, within the accuracy of our data, independent of
$t_w$. A similar time dependence has also been found in simulations of
spin-glasses~\cite{kisker96}, although in that case the exponent was
significantly smaller, and is thus not unusual for the aging dynamics.

The curves in Fig.~\ref{fig7}a are for the wave-vector $q=7.23$, the
location of the main peak in the static structure factor. In
Fig.~\ref{fig7}b we show the same type of correlation function for
different wave-vectors and $t_w=1000$. From the main figure we see that
the dependence of $C_k$ on $k$ is quite pronounced, in that e.g.  the
height of the plateau increases quickly with decreasing wave-vector.
Such a dependence can be understood at least qualitatively by recalling
that {\it in equilibrium} the wave-vector dependence of the
Edwards-Anderson parameter is very similar to a Gaussian, and that at
short times and long waiting times the time correlation functions do
not depend on $t_w$ (see Fig.~\ref{fig4}), i.e. show a
quasi-equilibrium behavior. From Fig.~\ref{fig7}b we see that a rough
estimate of $q_k$ is given by the value of $C_k(t_w+\tau,t_w)$ at
$\tau=20$ and we find that this quantity does indeed show a Gaussian
like dependence on $k$.

In the inset of the figure we show the same correlation functions in
a double logarithmic plot. From this graph we recognize that for all
wave-vectors the long time dependence of the functions are compatible 
with a power-law and that the exponent depends on the wave-vector in
that it decreases with decreasing $k$. 

Since at long times $C_k$ shows a power-law dependence it is not
possible to define a relaxation time. However, from the figure it
becomes clear that the relaxation of the system is much faster on small
length scales than on large ones. It is instructive to recall that for
a diffusion process the relaxation time depend on the wave-vector like
$k^{-2}$. A comparison of the curves for $k=3.0$ and $k=6.5$ shows that
the time needed to decay to 50\% of the initial value differs by almost
a factor of 100, thus much more than the factor of 4 expected for a
diffusive process, or the factor found in the relaxation dynamics of
supercooled liquids {\it in equilibrium}.  This shows that during the
aging process the relaxation at long times is indeed very different
from the one in supercooled liquids.

We now turn our attention to the time and waiting time dependence of
the two-time correlation functions at long times. Within mean-field
theory it is expected that for systems which show a discontinuous
dependence of the non-ergodicity parameter on temperature, which is the
case for the structural glass studied here, this dependence can be
written (in the limit of long waiting times) as

\begin{equation}
C(t_w+\tau,t_w)=C_{\rm ST}(\tau)+C_{\rm AG}\left(
\frac{h(t_w+\tau)}{h(t_w)}\right).
\label{eq10}
\end{equation}

Here $C_{\rm ST}(\tau)$ is the time dependence of $C(t_w+\tau,t_w)$ at
short times which is supposed to be independent of $t_w$ (in agreement
with our findings, see Fig.~\ref{fig4}) and to decay quickly to zero.
$C_{\rm AG}$ is a function whose form depends on $C(t_w+\tau,t_w)$ and
$h(t)$ is a monotonously increasing function of the argument. The
interesting point of this equation is that the whole $\tau$ and $t_w$
dependence of the aging regime enters only through the combination
$h(t_w+\tau)/h(t_w)$. Apart from Eq.~(\ref{eq10}), not much more is
known about the $\tau$ and $t_w$ dependence of $C(t_w+\tau,t_w)$.

In order to gain more insight into this dependence we investigate
whether the function $h(t)$ is independent of the correlation function
$C(t_w+\tau,t_w)$. The presence of such a dependence can, e.g., be
checked by plotting a time correlation function $C_{k'}(t_w+\tau,t_w)$
versus a different time correlation function, e.g.,
$C_k(t_w+\tau,t_w)$, i.e. by making a parametric plot with time
$\tau$ as the running parameter. If $h(t)$ is independent of the
correlation function considered, it is easy to see that in such a
parametric plot the curves corresponding to different waiting times
will fall on top of each other. (This holds true only if the time
dependence of $C_{ST}(\tau)$ can be neglected, i.e. at long times.)

\begin{figure}[h]
\psfig{file=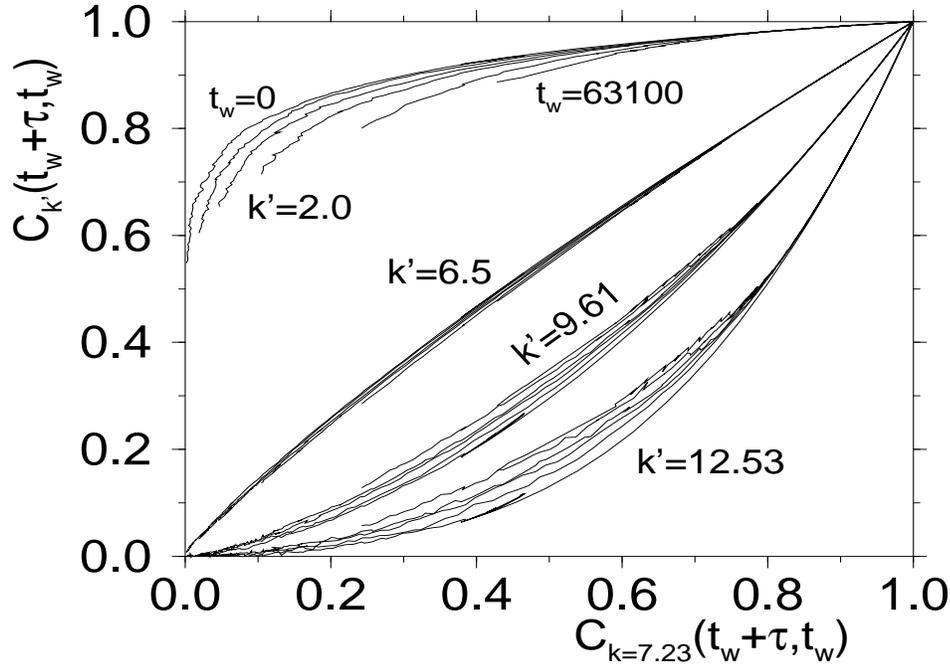,width=13cm,height=9.5cm}
\caption{Parametric plot between different correlation functions (see
axis labels). The different curves for given wave-vector correspond
to different values of the waiting time. $T_f=0.4$.}
\label{fig8}
\end{figure}

In Fig.~\ref{fig8} we show such parametric plots for the waiting times
$t_w=0$, 10, 100, 1000, 10000, and 63100 time units. The independent
variable (abscissa) is $C_k(t_w+\tau,t_w)$ for $k=7.23$ and the
dependent variables (the ordinate) are the same correlation function
for different values of $k$. Let us focus first on $k=2.0$. From the
graph we recognize that at short times (corresponding to large values
of $C_k$ and $C_{k'}$) the parametric plot is independent of $t_w$, as
it would be expected for the equilibrium case. However, for times at
which the correlation functions have fallen below their
Edwards-Anderson parameters a waiting time dependence is seen.
Qualitatively the same behavior is found for the other values of $k'$.
From this observation we hence conclude that if the Ansatz in
Eq.~(\ref{eq10}) is correct,  then the function $h(t)$ does depend
on the observable, i.e. in this case on the wave-vector. We also
mention that qualitatively the same results are obtained for the
different values of $T_f$.

\begin{figure}[h]
\psfig{file=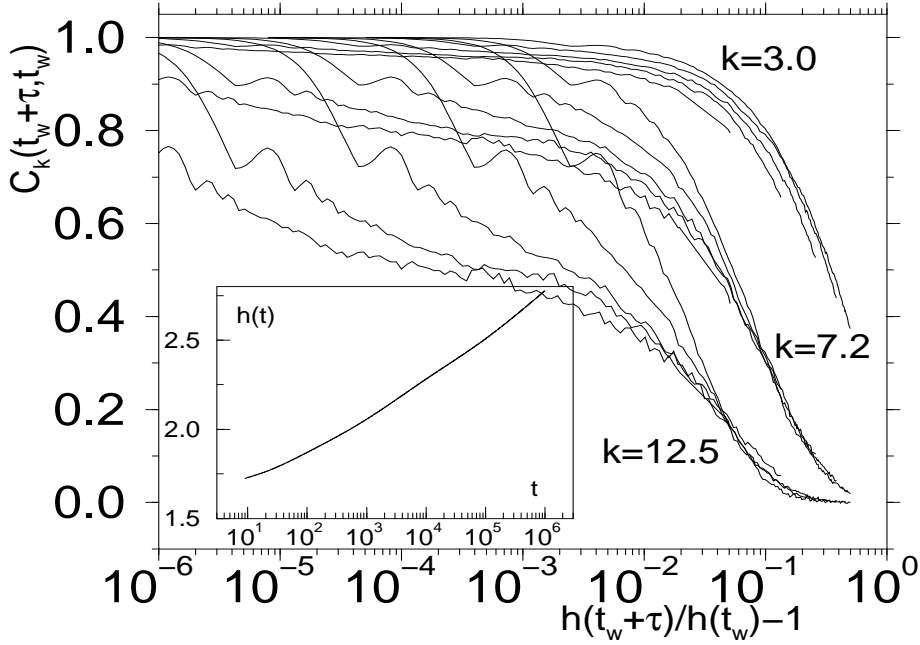,width=13cm,height=9.5cm}
\caption{Main figure: The two time correlation function as a function
of the variable $h(\tau+t_w)/h(t_w)-1$ for different wave-vectors and
waiting times (see text). The function $h(t)$ was chosen to make the
curves for $k=7.23$ collapse at long times and its time dependence is
shown in the inset.}
\label{fig9}
\end{figure}

The time dependence of the function $h(t)$ can be used to distinguish
between different theoretical models to describe the aging dynamics.
E.g. the droplet model of Fisher and Huse~\cite{fisherhuse} predicts
$h(t)$ to be of the form $h(t)=\log(t)$. In Ref.~\cite{mussel98} it was
argued that the present Lennard-Jones model showed this type of aging
dynamics, a conclusion which was not confirmed by the present
authors~\cite{kob98}.  The reason for this discrepancy remains still
unresolved. In order to settle this issue we have attempted to
determine the function $h(t)$ from our data without making reference to
any model, i.e.  functional form of $h(t)$. For this we assumed that at
long times the correlation functions are given by the second term in
Eq.~\ref{eq10}. Starting from the initial guess $h(t)=\log(t)$ we
plotted the correlation function versus $h(t_w+\tau)/h(t_w)$ for all
values of $t_w$, which gave at long times a clustering of the curves.
By iteratively bending $h(t)$ in a continuous way by small amounts and
subsequently making the scaling plot we attempted to generate a
collapse of the curves. The outcome of this procedure is shown in
Fig.~\ref{fig9} for the case $C_k$ with $k=7.23$. (For the sake of
clarity only the curves for $t_w=0$, 10, 100, 1000, 10000, and 63100
time units are shown, although more waiting times were considered to
determine the master curve. Also, in order to expand the abscissa we
have subtracted 1.0 from $h(t_w+\tau)/h(t_w)$ and plot its logarithm.).
From this figure we see that the it is indeed possible to obtain a
satisfactory scaling of the curves at long times. The function $h(t)$
which was obtained by the procedure described above is shown in the
inset. We see that to a first approximation it is close to a logarithm,
but that significant deviations are present.

Also included in the figure are the curves for different values of the
wave-vector. We see that in these cases the curves for the different
waiting times do not collapse nicely at long times. Thus this is
further evidence that the function $h(t)$ depends on the observable
considered. We also mention that although we have determined the
optimal shape of $h$ also for the other wave-vectors, it is difficult
to compare these different functions with each other.
 The problem is, that if $h(t)$ is
an optimal function that for example also  $h(t)^\alpha$
(with arbitrary $\alpha$) is an optimal function, i.e. there is no
unique representation of $h(t)$.

\subsection{Relaxation at low temperatures}

\begin{figure}[h]
\psfig{file=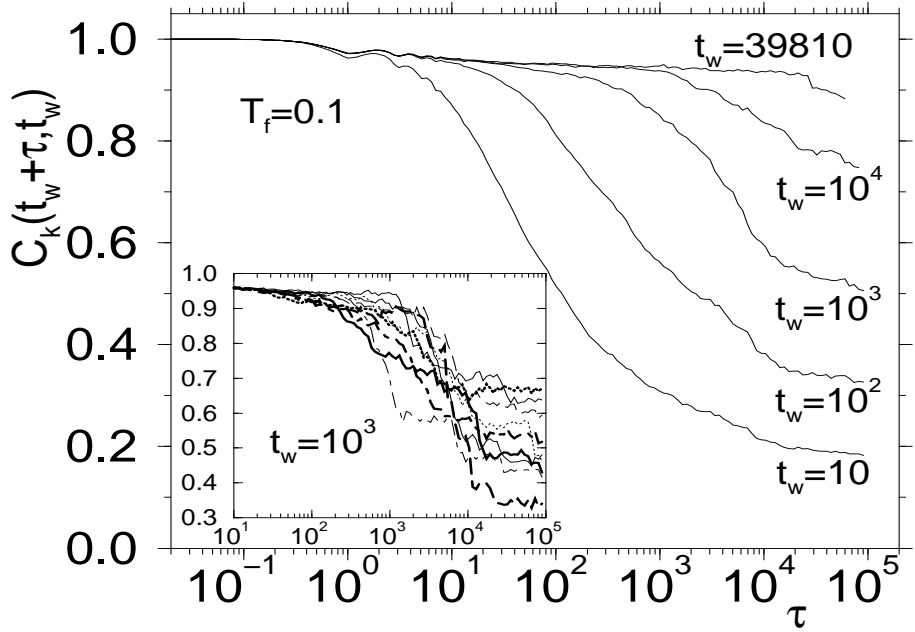,width=13cm,height=9.5cm}
\caption{Main figure: Time dependence of the correlation functions
$C_k(t_w+\tau,t_w)$, for the waiting times $t_w=10$, 100, 1000, 10000,
and 39810. $T_f=0.1$, $k=7.23$. Inset: The same correlation function
for $t_w=1000$ for the individual runs.}
\label{fig10}
\end{figure}

Essentially all the results discussed so far were obtained for a final
temperature $T_f=0.4$. This is only slightly (10\%) less that the mode
coupling critical temperature for the system,
$T_c=0.435$~\cite{kob_lj}. A markedly different behavior in the
relaxation is observed for  temperatures much lower than $T_c$, as
illustrated in figure ~\ref{fig10}, which shows the results for
$C_k(t_w+\tau,t_w)$  after a quench to $T_f=0.1$. The values of the
waiting times and of the wave-vector $k$ are the same as in figure
~\ref{fig4} ($T_f=0.4)$. At short times $\tau$, the dependences are
qualitatively similar for the two final temperatures. The only
difference is the increase in the plateau value when the temperature is
lowered.  This increase is easily understood from the fact that the
amplitude of the vibrational motion about an equilibrium position
decreases when $T$ decreases. From harmonic theory a linear dependence
of $1-q_{EA}$ on $T$ at low temperatures can be expected, and such a 
dependence is indeed compatible with our results.

For large values of $t_w$ and $\tau$, the behaviour for
$T_f=0.1$ is strikingly different from what was observed at $T_f=0.4$.
Instead of decaying rapidly to zero for  $\tau > t_w$, the correlators
level off and appear to display an additional plateau at a value of $C$
smaller than $q_{EA}$.

In order to get some insight into this surprising behavior, we have
studied separately the correlation functions obtained for various
samples prior to averaging. Looking at the data shown in fig.
~\ref{fig10} for some of the samples, it appears that the decay to this
second plateau is triggered by a large amplitude and rather sudden drop
of the correlation function, that, depending on the sample, takes place
typically $10^3$ to $10^4$ time units after the quench (see inset of
Fig.~\ref{fig10}). An analysis of the configurations shows that this
sudden drop is related to a very collective dynamical event, in which a
substantial fraction of the particles (10\%) move by a rather small
amount, typically 0.1-0.5$\sigma_{AA}$. Such motions can be understood from
the fact that a deep quench leaves the system in a highly stressed
configuration, so that the relaxation first proceeds through large
``earthquake like'' events, that release the local stress
significantly. Only at longer times the aging dynamics crosses over
to the very smooth process observed at higher values of $T_f$.

\begin{figure}[h]
\psfig{file=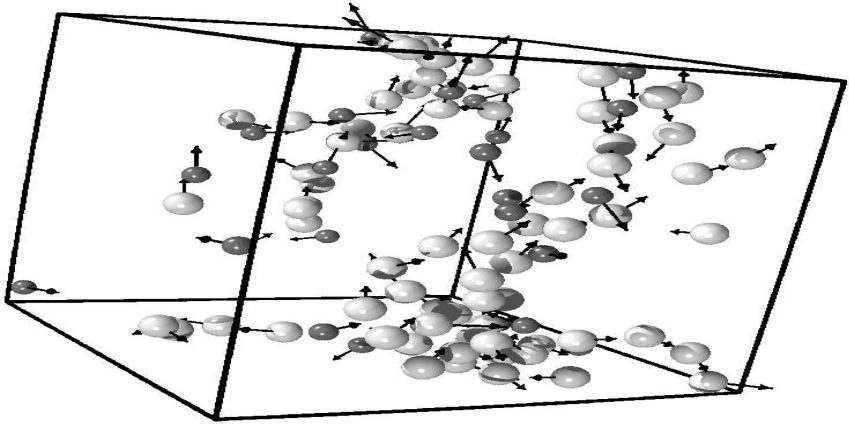,width=13cm,height=9.5cm}
\caption{Snapshot of the configuration just before ($\tau=5070$,
spheres) and just after ($\tau=7650$, tip of arrows) the large drop in
the time correlation function. In this event $C_k(t_w+\tau,t_w)$,
$k=7.23$, decayed from 0.79 to 0.52. $T_f=0.1$.}
\label{fig11}
\end{figure}

We emphasize that these events do not seem to be related to the hopping
like motion of the particles which is observed in the deeply
supercooled state {\it above} the glass transition
temperature~\cite{barrat90}. In that case only very few (2-3) particles
are involved and during the jump these particles move on the order of
one nearest neighbor distance. In contrast to this, the ``catastrophic
events'' during the aging are very collective (50-100 particles) and
the particles move only about 10-30 \% of the typical nearest neighbor
distance (as can be inferred from the self part of the van Hove
correlation function). A typical event is shown in Fig.~\ref{fig11}
where we show the particles just before the event (spheres) and their
location after the event (tip of arrows). From this figure we recognize that
in such an event the stress is released along a surface or even a line
through the sample and not through the motion of the particles in a
(bulk-like) three dimensional blob.

Obviously, the catastrophic events that  cause the decay of the
correlations in this situation are rather difficult to average over,
and a very large number of samples would be required to obtain a
reasonable statistical accuracy. Moreover, it is quite likely that
these events correspond to a transient, so that the second plateau we
observe is not really representative of the asymptotic behavior.
Interestingly, however, rather similar shapes of the time correlation
functions were observed by Bonn and coworkers \cite{bonn99} in their
dynamical light scattering studies of aging in clay (laponite)
suspensions. In that case the height of the second plateau was found to
steadily increase  with waiting time, confirming the transient nature
of the effect. 

We also mention that the occurence of these events are related
to the fact that the typical configuration at high temperature ($T_i=5.0$) are quite
different from the ones towards the system evolves to at $T_f$, thus
giving rise to large stresses. Since in {\it strong} 
glass formers (e.g. $SiO_2$) the structure has a much weaker temperature
dependance than in fragil eglassformers such as the present system, it can be expected that in the aging dynamics of the former no such
"catastrophic events" should be observed.

\section{Nonequilibrium response and fluctuation dissipation ratio}

\subsection{Definition and measurement of the nonequilibrium response}

One of the crucial points that was derived from the solution of the
dynamical equations in mean field models of spin glasses is the fact
that, in a nonequilibrium system, the Fluctuation Dissipation Theorem
(FDT) is violated in a nontrivial  way. We will first briefly recall
the main results from mean-field theories of spin glasses, and then
discuss how the violation of FDT can be quantified in our system.

Let us consider an observable $A$ whose normalized autocorrelation
function is denoted by $C$.  If $H$ denotes a field conjugate to $A$,
the response of $A$ to $H$ is defined as $R(t,t') = {\delta A(t) \over
\delta H(t')}$. In an equilibrium system, this response function  is
invariant under time translation, i.e.  $R(t,t')= R(t-t')$, and  is
related to the correlation function by the fluctuation dissipation
theorem, $R(t,t') = {1\over k_B T} {\partial C(t,t') \over \partial t'}$.
Out of equilibrium, this relation does not hold any more, and a
``Fluctuation Dissipation Ratio''  (FDR) $X(t,t')$ can be defined as

\begin{equation}
R(t,t') = {1\over k_B T} X(t,t') {\partial C(t,t') \over \partial t'} .
\label{eq17}
\end{equation}

Much attention has been devoted to the asymptotic behavior of this
fluctuation dissipation ratio. In particular, it has been shown
\cite{cugliandolo94} that in the asymptotic limit $t,t'\rightarrow
\infty$, the fluctuation dissipation ratio in mean field models of spin
glasses, as well as  in coarsening systems, becomes a non-singular
function of $C(t,t')$, i.e. $X(t,t')=x(C(t,t'))$.  A direct consequence
from this is that the more easily accessible integrated response

\begin{equation}
M(t_w+\tau,t_w)=   \int_{t_w}^{t_w+\tau} R(t_w+\tau, t) dt 
\label{eq18}
\end{equation}

can be written as a continuous function of $C$

\begin{equation}
M(t,t') = M(C) \int_C^1 x(c) dc\quad.
\label{eq19}
\end{equation}

From a practical point of view, the nontrivial consequence
of these statements is that a parametric plot of $M(t_w+\tau,t_w)$
versus $C(t_w+\tau,t_w)$ (with $\tau$ as the parameter) should, for
long times, converge towards a master curve, independent of the waiting
time. Such a behavior has been observed in a number of spin glass
simulations \cite{franzrieger,alvarez,marinari}.

In order to test whether the same property holds in structural glasses,
 we have to devise a way of measuring the response function
associated to $C_k(t,t')$, which is the only correlation function that
can be obtained with a reasonable accuracy in the aging regime.  The
procedure we use is the following.  A fictive ``charge" $\epsilon=\pm
1$ is assigned randomly to each particle. An additional term of the
form $\sum_j \epsilon_j V({\bf r_j})$, where $V({\bf r})  = V_0\cos
({\bf k}\cdot{\bf r})$ is a small ($V_0 <k_B T)$ external potential, is
then added to the  Hamiltonian. It is then easy to check that, {\it if
one averages over several realizations of the random charge
distribution}, the time-correlation function of the observable $A_k =
\sum_j \epsilon_j \exp(i{\bf k}\cdot r_j(t))$ is the incoherent
scattering function $C_k$. The procedure to generate the response
function associated to $C_k$ is thus straightforward: For a given
realization of the random charge distribution, the system is
equilibrated at a high temperature ($T_i=5.0$), and quenched at $t=0$ to
the desired final temperature $T_f$. The evolution is followed with the
field $V({\bf r})$ off until a waiting time $t_w$, then the field is
switched on and the response $A_k(t_w+\tau,t_w)$ is monitored. The same
procedure is repeated for several (7 to 10) realizations of the charge
distribution, in order to get the response function. The quantity we
obtain by this procedure is then an integrated response function
$M(t_w+\tau,t_w)$ (Eq.~\ref{eq18}), since

\begin{eqnarray}
\langle A_k(t_w+\tau, t_w)\rangle &=  &V_0 \int_{t_w}^{t_w+\tau} R(t_w+\tau, t) dt \\
& = & V_0 M(t_w+\tau,t_w) .
\end{eqnarray}

\begin{figure}[h]
\psfig{file=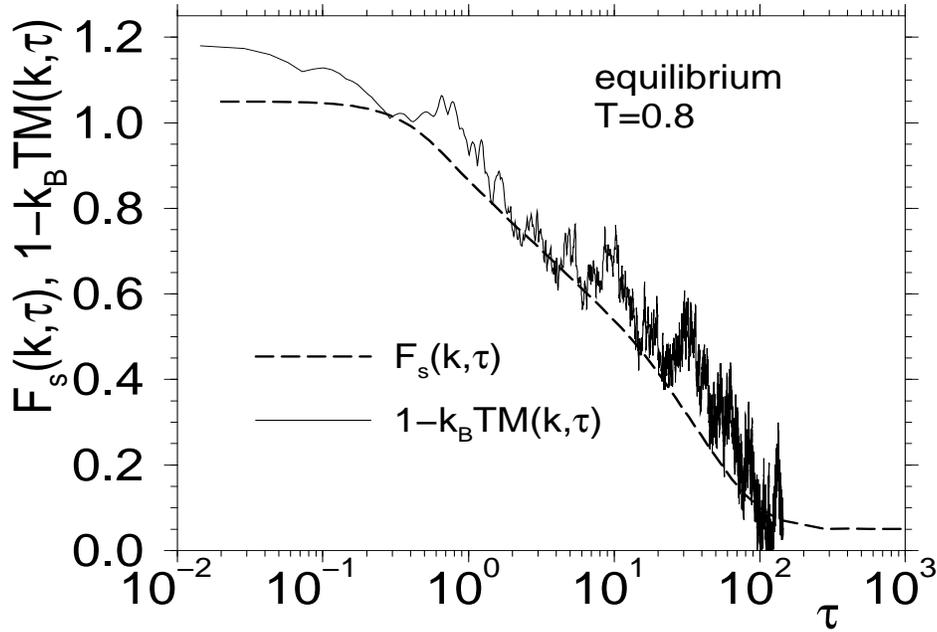,width=13cm,height=9.5cm}
\caption{Test of FDT at equilibrium. The dashed curve is the
intermediate scattering function $F_s(k,\tau)$ for $k=7.23$ and the
solid curve is the corrsponding response, i.e. $1-k_BTM(k,\tau)$. The
response data was obtained by averaging over 14 different runs.}
\label{fig12}
\end{figure}

We have checked that this procedure indeed yields the correct response
function by checking the fluctuation dissipation theorem in an
equilibrium system. In that case, a slightly different procedure was
used, in the sense that we monitored the decay of the response after
switching off the field. The fluctuation dissipation theorem implies
that this decay is directly proportional to the correlation function, a
result that is illustrated in fig. \ref{fig12}. From this figure we
see that the quality of the response data is significantly poorer than
the one for the correlation data. This is due to the fact that the
latter can be averaged over time origins and wave-vector directions,
hence improving drastically the statistical accuracy.

\subsection{Results for the response and the fluctuation dissipation ratio}

\begin{figure}[h]
\psfig{file=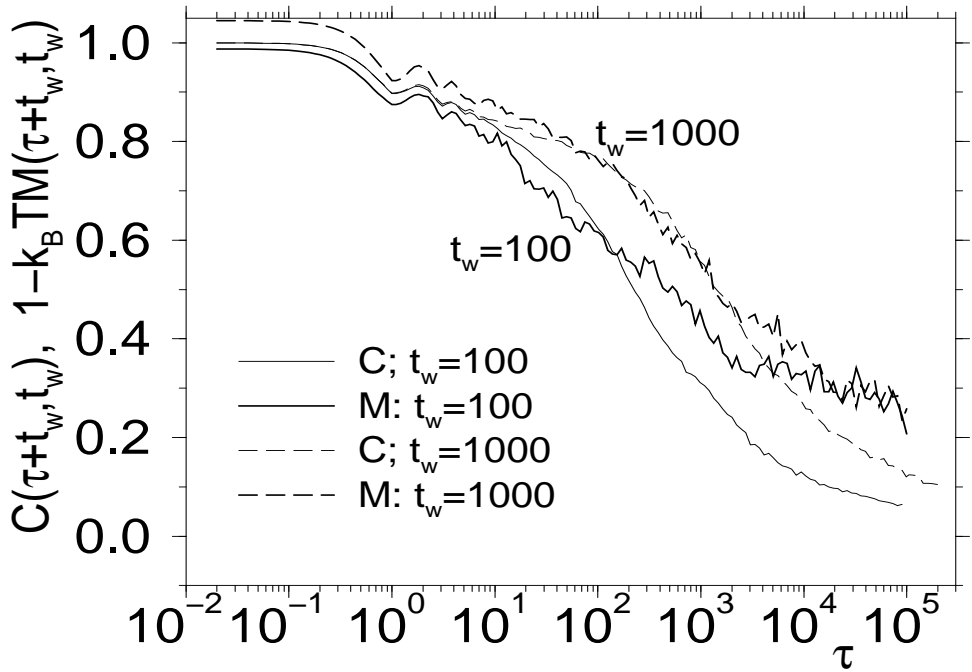,width=13cm,height=9.5cm}
\caption{Integrated response function $M$ and correlation function
for the out of equilibrium case.}
\label{fig13}
\end{figure}

For the nonequilibrium case typical results for the integrated response
function are shown in Fig.~\ref{fig13}, for $T_f=0.4$, $k=7.25$ and
two values of the waiting time. As expected, the response, like the
correlation, gets slower as the waiting time increases. As in the
equilibrium case, the quality of the response data is poorer than that
of the correlation data.  In that case, the average over wave-vector
directions in obtaining the correlation data probably explains this
difference. Despite this noise it can be seen from the figure that at
long times the curve for the response lies above the one for the
correlation function, i.e. that the response is smaller than expected
from the FDT.

As explained above, mean field theories of spin glasses suggest that
interesting information can be obtained from a parametric plot of
$M(t_w+\tau, t_w)$ versus $C(t_w+\tau, t_w)$ at fixed $t_w$.  Such a
plot is shown in figures \ref{fig14} (a-c) for three values of the
final temperature, $T_f=0.4$, $T_f=0.3$ and $T_f=0.1$.  The wave-vector
is $k=7.25$.  For each temperature, results obtained at different
waiting times are shown. In spite of the scatter in the data, the
figures are clearly compatible with the existence of a master curve
independent of the waiting time. Obviously, the quality of the data
does not allow a quantitative analysis of this master curve.  We can
nevertheless argue that two different regimes can be distinguished.
For high values of the correlation ($C_k$ close to unity), the
parametric plot is essentially a straight line with slope $-1$.  A
slope of $-1$ would be observed in an equilibrium system obeying FDT.
At lower values of the correlation, a clear deviation from this FDT
slope is observed. Although other fits are certainly possible, we can
describe the parametric plot as consisting of two straight lines, one
with slope $-1$ and one with a negative slope $-1< -m <0$. The reason
for choosing such a description and the corresponding interpretation
will be discussed in the next section.

A similar analysis was performed for another value of the wave-vector
($k= 3.0$). Very similar results are obtained, and the slope of $m$ of
the non-FDT part in the parametric plot seems to be independent of the
wave-vector.

Before discussing the results, we mention that the results for this
parametric plot are very sensitive to the linearity of the response.
Hence we always checked carefully that our results were independent of
the amplitude of the applied field.  In practice, applying a stronger
field does not affect the FDT part of the parametric plot, but tends to
affect strongly the second part, in that it leads to a
decrease in the parameter $m$ that describes the non-FDT part of the
curve.

\subsection{Discussion}
\label{secVC}

The importance of the fluctuation dissipation ratio  $x(C)$ in glassy
systems was first recognized in the context of the mean field theory of
spin glasses \cite{cugliandolo94}, when it appeared that  this FDR was
intimately related to the nature of ergodicity breaking in the system.
The nature of this relationship was recently clarified by Franz {\it et al.} \cite{franz98}, who showed that the asymptotic value of the
FDR could be quite generally expressed in terms of the probability
distribution $P(q)$ of overlaps between replicas introduced by Parisi
\cite{fisherhertz} through

\begin{equation}
x(C)= \int_0^C P(q) dq.
\label{eq20} 
\end{equation}

This relation between a purely static quantity, $P(q)$, and a dynamical
one, $x(C)$,  is  extremely powerful, since it implies that a
nontrivial information on the nature of phase space may be encoded in
the nonequilibrium time dependent behavior.

If we accept equation (\ref{eq20}), a small number of possible scenarios
are documented in the literature \cite{parisi_ricci}. For systems with
continuous step replica symmetry breaking, $P(q)$ is a continuous function,
and so is $x(C)$. For systems with one step replica symmetry breaking
like $p$-spins systems ($p>3$), $P(q)$ consists of two $\delta$-functions
at $q=0$ and $q=q_{EA}$, so that $x(C)=1$ for $C>q_{EA}$ and $x(C)=m<1$
for $C<q_{EA}$. Finally simple coarsening systems, like the Ising
system below its ferromagnetic transition, have an essentially trivial
$P(q)$, $P(q) = \delta (q-M^2)$, where $M=M(T)$ is the magnetization.
Therefore,  in that case the FDR is  1 if $1> C > M^2$, and 0 if $ M^2
> C$ \cite{letitia,abarrat,lberthier}.

\begin{figure}[h]
\psfig{file=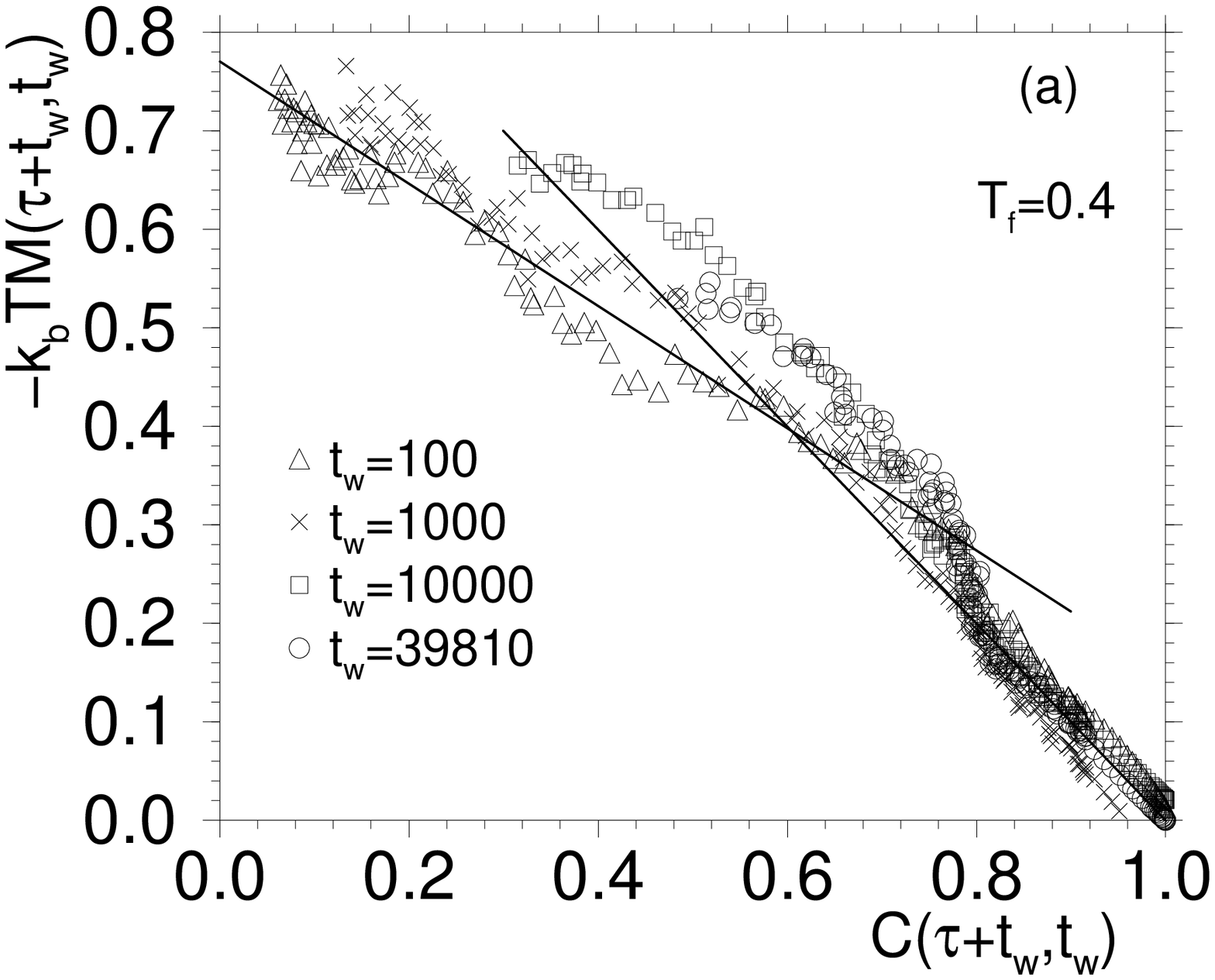,width=13cm,height=9.5cm}
\psfig{file=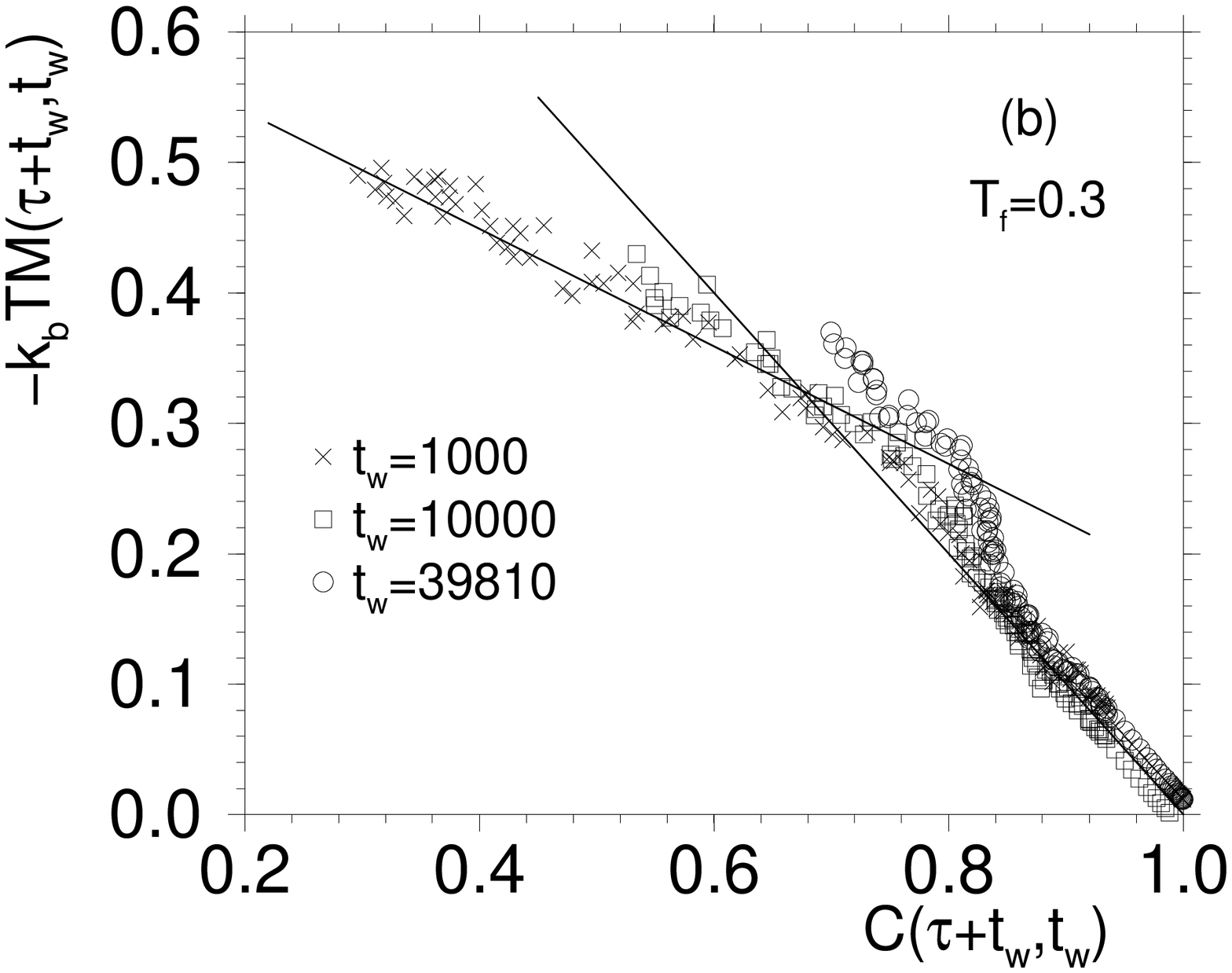,width=13cm,height=9.5cm}
\psfig{file=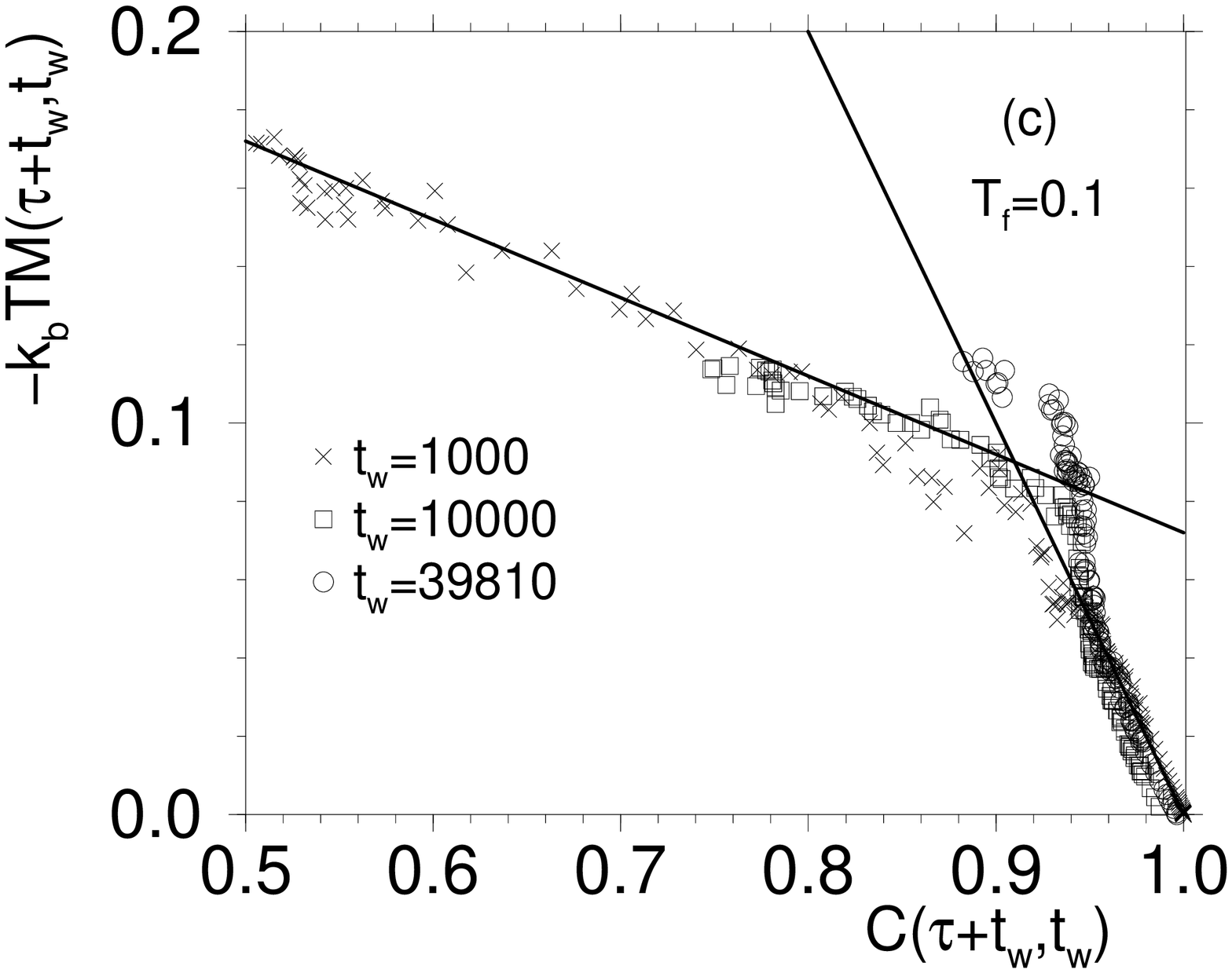,width=13cm,height=9.5cm}
\caption{Parametric plots a) $T_f=0.4$  b) $T_f=0.3$ c)$T_f=0.1$}
\label{fig14}
\end{figure}

Within this theoretical framework, we have to find which scenario is
compatible with the results presented in the last section.
From Figs. \ref{fig14}, it  would appear that the most likely
scenario is that of a system with one step replica symmetry breaking,
for which $x(C)$ is a stepwise constant function.  In that case the
parametric plot consists in two straight lines, one with slope $-1$ and
one with slope $ -1<-m<0 $ (bold straight lines).

Obviously the asymptotic nature of the results can be questioned.
However, the fact that we obtained an essentially $t_w$ independent
plot is already an indication that we are approaching the asymptotic
limit.  It could also be that preasymptotic effects cancel out when the
parametric plot is used. Indeed,  such a plot is even obtained for
$T_f=0.1$, where we have seen that strong ``catastrophic'' events
influence the relaxation.  Finally, we mention that in systems in which
preasymptotic effects have been observed and studied (mostly domain
growth models) \cite{parisi_ricci,lberthier}, they very clearly show up
in the parametric plot as a $t_w$ dependence of the crossover region between
the FDT part and the non-FDT region. The overall shape of the
parametric plot is not affected much.  Such a dependence could then
explain the fact that in our data, the crossover between the FDT part
and the non-FDT part takes place at a value of $C$ {\it smaller} than
the plateau value $q_{EA}$ in the correlation function, a feature which
contradicts theoretical expectations. The tendency of the curves to
``overshoot'' in the crossover region,  perceptible in Fig.
\ref{fig14}, is reminiscent of the observations made in reference
\cite{parisi_ricci},  and could also be a transient effect. In that
paper it was argued that, e.g., in a coarsening system with some
defects the de-pinning of a domain wall from a defect will give rise
to a strong enhancement of the response just after the event. Since
 the ``catastrophic'' events correspond to a violent and, most likely,
pre-asymptotic release of the stress, it is reasonable to assume that
also in this case the response will be larger than expected.
 
If we accept that our parametric plots correspond to the one step
replica symmetry breaking case, we can extract from the data an
estimate for the slope $-m$ of the non-FDT part.  The resulting slopes
are given by:  $T_f=0.4$, $m=0.62 \pm 0.05$; $T_f=0.3$, $m=0.45\pm
0.05$; $T_f=0.1$, $m=0.2\pm 0.1$. As mentioned above, these results
appear to be independent of the wave-vector. Within the accuracy of our
data, they are compatible with a linear dependence of $m$ on $T_f$,
quite similar to that found by Parisi~\cite{parisi97} for a soft-sphere
system. We also note that the value found for $T_f=0.4$ is compatible
with the result $m=0.57$ that was extracted from a scaling analysis of
the correlation functions alone.

It is interesting to analyze our results for $m$ using the ``effective
temperature'' concept introduced in reference \cite{CKP}. (For a
different approach to the effective temperature see the papers of 
Nieuwenhuizen~\cite{nieuwenhuizen98}).  In this
approach, an effective ``fluctuation dissipation temperature'' $T_{eff}$
is defined as $T_{eff} = T/m$, where $T$ is the actual external
temperature.  Crudely speaking, the ``fast'' degrees of freedom (those
that correspond to the rapid decay of $C$ to its plateau value) are at
equilibrium with the thermostat at temperature $T$, while the slow
degrees of freedom, which govern the aging behavior, are at equilibrium
at a higher temperature. Then the concept of effective temperature can
be thought as a rationalization of the older ``fictive temperature''
concept \cite{brawer}.  Within this interpretation, a linear dependence
of $m(T_f)$ on $T_f$ corresponds to a constant effective temperature.
It is therefore quite natural to expect first that $m(T_f)$ should be
equal to $T_f/T_c$, where $T_c$ is the mode coupling critical
temperature, since systems cooled below $T_c$ fall out of equilibrium
at $T_c$.  This  result was proposed by Parisi \cite{parisi97} on the
basis of his simulations in a soft sphere system. Our results do not
corroborate such a simple interpretation. The ``effective temperature''
in our aging system is substantially larger than $T_c$, typically
$T_{eff}\simeq 0.7$.  Our results here are  similar those  of Alvarez
{\it et al.} \cite{alvarez96} for the $p$-spin model in that these
authors found for a temperature a bit below $T_c$ a value of $m$ which
is significantly smaller than 1.  Indeed, the ``ideal'' result
$m(T_f)=T_f/T_c$ can only be expected to hold for systems that would be
cooled infinitely slowly, so that they would remain in equilibrium down
to $T_c$, and for which ergodicity restoring hopping processes can be
completely neglected. For a system that is cooled with a finite cooling
rate, it is not surprising to find an effective temperature which is
above $T_c$.

\section{Conclusions}

In this paper, we have presented a detailed numerical study of the out
of equilibrium relaxation (``aging'') of a simple glass forming system.
The time scales we have investigated are those allowed by the current
possibilities of Molecular Dynamics simulations, typically $10^{-8}-10^{-7}$s.
We will now try to briefly summarize the main conclusions that can be
inferred from these numerical results.

First, it appears that static quantities are, on these time scales,
only very weakly dependent of time. By static (or ``one time'')
quantities we mean, as usual, those quantities that can be obtained
from the knowledge of a single configuration of the system.  In
particular, we have shown that the pair correlation functions (and
hence all the derived quantities like the energy or the pressure)
appear to equilibrate very quickly after the quench.  These one-time
quantities show some sensitivity  to the quench history and to the
final temperature and we find that to a first approximation they
correspond to configurations which are very close to the critical
surface of (equilibrium) mode-coupling theory. Hence we have evidence
that for this model and within the time span of the simulation the
system is not able to penetrate this surface and hence remains in the
liquid like part of configuration space.

In contrast to this weak dependence, the two-time  correlation
functions are very sensitive to the aging time. When the quench
temperature is relatively high (i.e. still close to the mode-coupling
critical temperature that was identified for our system), the waiting
time dependence of these functions can be described in rather simple
terms. Functions obtained for various $t_w$ can be rescaled on a master
curve, and an analysis of  this curve yields results that are
compatible with the predictions of mean field (``mode-coupling'')
theories  of the glass transition formulated for the ``$p$-spin'' spin
glass models.

At lower final temperatures, the behaviour of the two-time correlation
functions is much more complex. The relaxation is largely dominated by
``catastrophic events'' that involve a small, but collective,
displacement of a large number of particles.  These infrequent and
large events are very difficult to average over, so that the
statistical quality of the data for the correlation functions is rather
poor.  Due to the limited time range of the simulations, we do not know
whether these events are transients effects releasing some initial
stresses due to the quench, or constitute  a genuine characteristic of
low temperature relaxation.

The flexibility of MD simulations allows an independent measurement of
the two-time response functions, which at equilibrium would be related
to the correlation functions through the fluctuation dissipation
theorem (FDT).  A parametric representation of the response versus the
correlation allows to clearly distinguish between a ``quasi-equilibrium''
region in which the FDT holds, and a second regime in which it is
violated.  An essential  feature is that this parametric plot does not
(or, at most, very weakly) depend on waiting time, so that the ``FDT'' and
``non-FDT'' regimes do not correspond to {\it time windows} but rather to
{\it correlation} windows.  This remarkable feature, predicted by mean
field theories of spin glasses to hold in the asymptotic limit, is
observed here for structural glasses at finite times.  This is likely
related to the  formal similarities between the mode-coupling theories
of structural glasses, which are known to describe quite well the
equilibrium behaviour of our system,  and these mean field models of
spin glasses.  As  the response/correlation relationship observed for
our model resembles the one observed in spin systems with one step
replica symmetry breaking, one may speculate that the phase space
structures in both systems are also similar. If this is the case, a
quite appealing scenario for the glass transition can be devised, that
reconciles some of  the old ideas of the Adams-Gibbs approach with the
more modern mode-coupling scenario, as discussed in detail in reference
\cite{mezardparisi}.

Unfortunately, we presently do not have  at our disposal a quantitative
theory of nonequilibrium behaviour in structural glasses.  The
theoretical framework currently provided by mean-field theories of spin
glasses can be considered as a schematic model, and hence does not allow
comparisons beyond the qualitative level. At this level, we find that
the general behaviour of our system is in quite good agreement with
this schematic model.  The agreement (and disagreements) are somewhat
reminiscent of what is observed when comparing equilibrium data with
predictions of schematic mode-coupling theories.

Finally, it is natural to inquire about  the relevance of a work on aging
phenomena based on a method that is limited to time scales smaller than
$10^{-7}$s to the aging effects observed on much longer
time scales.  The general picture we obtain provides, however, some
support for a scenario \cite{mezardparisi}
 that may  be of general relevance for much
longer time scales.  In this scenario, the critical temperature $T_c$
of mode coupling theory is associated to the dynamical freezing
temperature of mean field spin models. In a purely mean field (or, in
other words, ideal mode-coupling) situation, the system would be out of
equilibrium at all temperatures below $T_c$. Of course, real systems
can still be equilibrated below $T_c$, as ``hopping processes'' allow a
relaxation that is not described by mode-coupling theory (at least in
its ideal version). Hence in real experiments (or in our system, if we
could allow for longer simulation times) these systems will show only
interrupted aging, i.e. aging for short waiting times and no aging at
long waiting times. However, we have shown that if the system was
artificially driven into a nonequilibrium situation, its ``short time
aging'' can be described reasonably well within the framework of
mean-field/mode-coupling theories. We may here draw a parallel with the
fact that, in equilibrium,  below $T_c$, when the  $\alpha$ (terminal) relaxation becomes
dominated by hopping processes, the $\beta$ (intermediate times)
relaxation is still very well accounted for by ideal mode coupling
theory \cite{mct,mct_exp}. It is then quite tempting to speculate that
for any ``fragile'' system (i.e. a system in which hopping processes are
not too strong, so that mode coupling theory accounts well for the
dynamics close to $T_c$) quenched below its glass transition, the
mean-field/mode-coupling description will be relevant for a large
fraction of the aging process. This should be particularly relevant in
systems where hopping processes are known to be weak, like colloidal
suspensions.

\acknowledgments

We acknoweledge fruitful discussions with 
L. Cugliandolo, J. Kurchan and A. Latz. This work was supported
by the P\^ole Scientifique de Mod\'elisation Num\'erique at ENS-Lyon, 
and by the Deutsche Forschungsgemeinschaft under SFB262.

\end{document}